\documentclass{ieeetmlcn}
\usepackage[nobreak]{cite}
\usepackage{amsmath,amssymb,amsfonts}
\usepackage{graphicx,color}
\usepackage{textcomp}
\usepackage{xcolor}
\usepackage{hyperref}
\hypersetup{hidelinks}
\usepackage{algorithm}
\usepackage{algpseudocode}
\usepackage{cleveref}
\usepackage{float}
\usepackage{url}
\makeatletter
\let\MYcaption\@makecaption
\makeatother
\usepackage{subcaption}
\makeatletter
\let\@makecaption\MYcaption
\makeatother
\usepackage{multirow}
\usepackage{scalefnt}
\usepackage{booktabs}
\usepackage{bm}
\usepackage{comment}

\usepackage{mathtools}

\usepackage{optidef}

\crefname{figure}{Fig.}{Figs.}
\crefname{table}{Table}{Tables}
\crefname{algorithm}{Algorithm}{Algorithms}
\crefname{section}{Section}{Sections}

\algdef{SE}[SUBALG]{Indent}{EndIndent}{}{\algorithmicend\ }
\algtext*{Indent}
\algtext*{EndIndent}
\algrenewcommand\algorithmicrequire{\textbf{Input:}}
\algrenewcommand\algorithmicensure{\textbf{Output:}}

\def\BibTeX{{\rm B\kern-.05em{\sc i\kern-.025em b}\kern-.08em
    T\kern-.1667em\lower.7ex\hbox{E}\kern-.125emX}}
\AtBeginDocument{\definecolor{tmlcncolor}{cmyk}{0.93,0.59,0.15,0.02}\definecolor{NavyBlue}{RGB}{0,86,125}}

\def\OJlogo{\vspace{-4pt}\includegraphics[height=18pt]{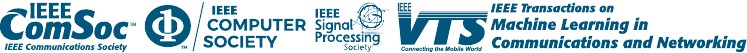}}
\def\seclogo{\vspace{10pt}\includegraphics[height=18pt]{new_logo}}

\def\authorrefmark#1{\ensuremath{^{\textbf{#1}}}}

\begin{document}
\receiveddate{December 19, 2022}
\reviseddate{April 16, 2023, July 4, 2023, and August 17, 2023}
% \accepteddate{XX Month, XXXX}
% \publisheddate{XX Month, XXXX}
\currentdate{August 17, 2023}
% \doiinfo{TMLCN.2022.1234567}

\markboth{Point Cloud-based Proactive Link Quality Prediction for Millimeter-wave Communications}{Ohta et al.}

\title{Point Cloud-based\\Proactive Link Quality Prediction\\for Millimeter-wave Communications}

\author{
Shoki~Ohta\authorrefmark{1}, Student~Member,~IEEE,
Takayuki~Nishio\authorrefmark{1},
Senior~Member,~IEEE,
Riichi~Kudo\authorrefmark{2},
Member,~IEEE,
Kahoko~Takahashi\authorrefmark{2},
and Hisashi~Nagata\authorrefmark{2}
}
\affil{School of Engineering, Tokyo Institute of Technology, Tokyo 152-8550, Japan}
\affil{NTT Network Innovation Laboratories, NTT Corporation, Yokosuka 239-0847, Japan}
\corresp{Corresponding author: Takayuki~Nishio (email: \href{mailto:nishio@ict.e.titech.ac.jp}{nishio@ict.e.titech.ac.jp}).}
\authornote{This work was supported in part by JSPS KAKENHI Grant Number JP22H03575.\\This article was presented in part at the 95th IEEE Vehicular Technology Conference (VTC2022-Spring).}

\begin{abstract}
This study demonstrates the feasibility of point cloud-based proactive link quality prediction for millimeter-wave (mmWave) communications.
Previous studies have proposed machine learning-based  methods to predict received signal strength for future time periods using time series of depth images to mitigate the line-of-sight (LOS) path blockage by pedestrians in mmWave communication. 
However, these image-based methods have limited applicability due to privacy concerns as camera images may contain sensitive information. 
This study proposes a point cloud-based method for mmWave link quality prediction and demonstrates its feasibility through experiments.
Point clouds represent three-dimensional (3D) spaces as a set of points and are sparser and less likely to contain sensitive information than camera images.
Additionally, point clouds provide 3D position and motion information, which is necessary for understanding the radio propagation environment involving pedestrians.
This study designs the mmWave link quality prediction method and conducts realistic indoor experiments, where the link quality fluctuates significantly due to human blockage, using commercially available IEEE 802.11ad-based 60\,GHz wireless LAN devices and Kinect v2 RGB-D camera and Velodyne VLP-16 light detection and ranging (LiDAR) for point cloud acquisition.
The experimental results showed that our proposed method can predict future large attenuation of mmWave received signal strength and throughput induced by the LOS path blockage by pedestrians with comparable or superior accuracy to image-based prediction methods. 
Hence, our point cloud-based method can serve as a viable alternative to image-based methods.
\end{abstract}

\begin{IEEEkeywords}
LiDAR, link quality prediction, machine learning, millimeter-wave communication, point cloud
\end{IEEEkeywords}


\maketitle

\section{INTRODUCTION}\label{s:intro}
\IEEEPARstart{W}{ith} the rapid expansion of wireless communication applications, the microwave frequency band is strained and the utilization of higher frequency bands, such as millimeter-wave (mmWave) is underway.
mmWave communication is crucial for extremely high transmission rate in the fifth-generation (5G) mobile communication system and wireless local area network (WLAN) standard IEEE 802.11ad/ay because mmWave communication can provide wide bandwidth~\cite{uwa2020access, hong2021role, nitsche2014commag, hansen2011wigig}.
This high transmission rate enables applications that require significant amounts of traffic, such as virtual and augmented realities (VR/AR), environment sensing, and ultra-high-definition (UHD) video streaming.
Thus, mmWave communications greatly increase the possibilities of wireless communications and are expected to have a variety of applications.

Despite its wide bandwidth, mmWave communication has technical challenges, such as sensitivity to line-of-sight (LOS) path blockage, radio directivity, significant path loss, and narrow beamwidth, owing to its short wavelengths.
The link quality significantly deteriorates when the mmWave LOS path is blocked by a human body or vehicle~\cite{collonge:twc2004influence}.
mmWave communications are expected to be used for indoor and dense urban environments, such as VR/AR applications in private residences, environment sensing and equipment control in factories, and UHD video streaming at event venues.
Such indoor or dense urban environments are common in residential or industrial spaces, and the mmWave LOS path blocked by human bodies or robots occurs frequently.
When mmWave is used under these conditions, communication disconnections occur frequently and the average throughput significantly decreases compared with LOS communication.
Therefore, it is effective to predict the future wireless communication environment and adaptively control communications in order to fully utilize the high transmission rate of mmWave communications.

Traditional methods such as empirical and stochastic analysis to provide stochastic prediction models~\cite{collonge:twc2004influence, stochastic} and time series forecasting~\cite{wsn, mesh} have been investigated for mmWave communications. 
However, accurately predicting future link quality in mmWave communication, where the link quality changes sharply due to human blockage, has been challenging. 
This is because human blockage occurs non-periodically, and there are no apparent signs of deterioration in link quality until the negative effects appear.

To address this challenge, computer vision-aided (CV-aided) mmWave communications have been proposed and are gaining a lot of attention~\cite{nishio:csm2021when, nishio:jsac2019proactive, wu:wcnc2022lidar}.
The use of camera image and computer vision (CV) techniques, including machine learning (ML) in mmWave communications, enables accurate prediction of link quality such as received signal strength and throughput. 
ML algorithms can learn to accurately map the relationship between the camera images and link quality by analyzing camera images of the mmWave propagation environment, including the geometry and dynamics of obstacles.
Compared with traditional methods, which often rely on empirical channel models or time-series forecasting, the combination of camera images and ML provides a more deterministic and accurate approach to predicting future link quality.
Proactive communication controls such as transmission power control, base station handover, beamforming, frequency switching, and intelligent reflecting surface (IRS) control~\cite{gong:irs} can be performed to mitigate mmWave LOS blockage effects based on the accurate and deterministic link prediction.
Our previous work demonstrated that a deep learning-based method can predict mmWave received signal strength 500\,ms ahead from depth camera images~\cite{nishio:jsac2019proactive}.

However, images may contain confidential information, particularly in private residences, offices, and factories.
This property limits the application scenarios of the existing CV-aided mmWave communication systems that leverage cameras. 
Therefore, alternative sources of information on the mmWave communications environment are required.

This study proposes a link quality prediction method using point clouds as an alternative to images.
A point cloud represents three-dimensional (3D) space as a set of points and can be obtained by light detection and ranging (LiDAR), or depth cameras. 
LiDAR estimates the distance to objects by measuring the time difference between the emission of light and the arrival of the reflected light~\cite{yilmaz2006object}.
Compared with images, point clouds are sparse and less likely to contain private information~\cite{gunter2020privacy, rodrigues2021laflector}.
Owing to privacy concerns, LiDAR is increasingly being installed in place of cameras for sensing.
Point clouds have many applications such as robot operations~\cite{ros}, autonomous driving~\cite{geiger2012kitti, caesar2020nuscenes}, and digital twin~\cite{deng2021systematic, digitaltwin}.
Wireless communications are expected to increase in value by integrating with these fields.
Further, point clouds acquired from LiDAR are superior to images in terms of 3D position accuracy and lighting robustness.
Cameras may not be able to observe objects at distant locations or accurately measure distances.
Point clouds obtained from LiDAR have more detailed coordinate information of 3D space than images because the surface of an object can be accurately obtained as 3D information~\cite{villani2018survey}.
Cameras are susceptible to sun glare, such as direct light and backlight~\cite{yoneda:iv2021sun}, and using them in the dark is difficult.
Our LiDAR point cloud-based system can operate the link quality prediction system without the influence of sunlight or lighting.
Therefore, point clouds can be used as an alternative feature to images in predicting link quality.

The main objective of this paper is to showcase the possibility of predicting mmWave link quality using point clouds, for which we propose a prediction method based on ML. 
As there is currently no established method for link quality prediction using point clouds, similar to image-based methods, we rely on ML which has proven to perform well in various point cloud and computer vision tasks.
Previous studies~\cite{wu:wcnc2022lidar, marasinghe:gcw2021lidar} showed that the link state, i.e., LOS or non-LOS (NLOS), can be predicted from the point cloud.
However, the quantitative prediction of the future received signal strength or throughput (e.g. 500\,ms or 1000\,ms ahead), which enables fine-grade link control but is a more challenging task than classifying LOS or NLOS, was out of scope.
Furthermore, the conventional image-based prediction method utilizing deep learning~\cite{nishio:jsac2019proactive} cannot be applied to point cloud-based prediction owing to the large data domain gap between point clouds and images. 
Therefore, we construct a preprocessing method of point clouds suitable for the link quality prediction, which transforms point clouds into a different representation of 3D space, voxel grids.
We then selected regression ML algorithms for link quality prediction that can be applied to voxel grids.

This study demonstrates the feasibility of the point cloud-based link quality prediction by conducting experiments in an environment closer to practical environments compared with existing study~\cite{nishio:jsac2019proactive}.
Commercially available IEEE 802.11ad-compliant devices were used for the access point (AP) and the station (STA) during experiments. 
Two numerical indicators with slightly different characteristics, received signal strength indicator (RSSI) and throughput, were used to evaluate link quality. 
Additionally, two types of point clouds with different properties acquired with different devices, LiDAR and depth cameras, were utilized.

The contributions of this paper are summarized as follows:
\begin{itemize}
    \item We have demonstrated the feasibility of proactive mmWave link quality prediction using point clouds. 
    The experiments were conducted in indoor environments, where the link quality fluctuates significantly due to human blockage, using commercially available IEEE 802.11ad-based 60\,GHz wireless LAN devices and Kinect v2 RGB-D camera and Velodyne VLP-16 LiDAR for point cloud acquisition. 
    The experimental results show that our point cloud-based method can model the relationship between spatial variation and mmWave RSSI or throughput variations through ML algorithms and predict future RSSI or throughput up to 1000\,ms in advance, without dependence on point cloud acquisition devices.
    \item We have formulated mmWave link quality prediction from point clouds as a regression task, and developed a novel method for performing such predictions. 
    Our method involves a series of preprocessing steps, including removing excess areas and outliers, downsampling, and converting point clouds into voxel grids with corresponding labels. 
    This allows us to quantitatively and deterministically predict link quality through supervised learning. 
    We examined three supervised learning models, a neural network (NN) with 3D convolution layers, a NN with 3D convolutional long short-term memory (LSTM) layers, and gradient boosting decision tree (GBDT), to learn the mapping from voxel grids to link quality.
    \item We compared the proposed point cloud-based link quality prediction method, a time series link quality-based method, and a previous depth image-based method~\cite{nishio:jsac2019proactive} through experiments.
    Our point cloud-based method can predict RSSI with an error of less than 3.99\,dB and throughput with an error of 0.313\,Gbit/s up to 1000\,ms ahead, which can be compared to or outperforms the image-based method.
    In contrast, the time series-based method cannot make accurate predictions.
\end{itemize}

This paper expands upon our conference paper~\cite{ohta:vtc2022mill}.
We have conducted new experiments using LiDAR and provided a more in-depth evaluation of both the proposed method and the image-based method. 
Additionally, we have enhanced the explanation of the proposed method and included comprehensive discussions on related works and future research directions.

The remainder of this paper is organized as follows.
\cref{s:related_works} describes related works on mmWave link quality prediction and point clouds application for wireless communication.
\cref{s:proposedmethod} describes the system model, problem formulation, preprocessing method, and prediction methods of our point cloud-based mmWave link quality prediction.
In Sections~IV and V, our proposed method is evaluated through experiments using depth camera point clouds and received signal strength datasets, and LiDAR point clouds and throughput datasets, respectively.
Section~VI discusses remaining challenges and future research directions.
Section~VII concludes this paper.
\section{RELATED WORKS}\label{s:related_works}
This section summarizes existing research on mmWave link quality prediction and applications of point clouds for wireless communication.
As mentioned in \cref{s:intro}, the link quality prediction task is critical in the proactive control of mmWave communications.
Therefore, various methods have been proposed, including those specialized for indoor and outdoor environments, as well as methods applying computer vision techniques using images and point clouds.
\cref{tab:rlwk1} shows the related works on link quality prediction and applications of the point cloud.
Our work is orthogonal to these studies and increases the potential for mmWave communication.
\cref{tab:rlwk2} provides a more detailed summary of link quality prediction studies~\cite{nishio:jsac2019proactive, wu:wcnc2022lidar, marasinghe:gcw2021lidar, klautau:wcl2019lidar, zhang:infocom2022vision, asano:apcws2021high}, which are more closely related to our proposed method. In the following, we discuss the differences between these studies and ours.

\begin{table*}[t!]
\centering
\caption{Summary of related works on link quality prediction and applications of point cloud.}
\begin{tabular}{|c|c|c|c|c|c|c|c|c|c|c|c|}
\hline
Existing works              & \cite{nishio:jsac2019proactive} & \cite{wu:wcnc2022lidar} & \cite{marasinghe:gcw2021lidar} & \cite{klautau:wcl2019lidar} & \cite{zhang:infocom2022vision} & \cite{asano:apcws2021high} & \cite{egi:access2019machine} & \cite{jan:lwc2016evaluation} & \cite{jan:tap2016indoor} & \cite{stephan:eucap2018increased} & \textbf{Ours} \\ \hline 
Link quality prediction     & $\sqrt{}$                       & $\sqrt{}$               & $\sqrt{}$                      & $\sqrt{}$                   & $\sqrt{}$                      & $\sqrt{}$                  & $\sqrt{}$                    & $\sqrt{}$                    &                          &                                   & $\sqrt{}$     \\ \hline
mmWave                      & $\sqrt{}$                       & $\sqrt{}$               & $\sqrt{}$                      & $\sqrt{}$                   & $\sqrt{}$                      & $\sqrt{}$                  &                              &                              & $\sqrt{}$                & $\sqrt{}$                         & $\sqrt{}$     \\ \hline
Point cloud                 &                                 & $\sqrt{}$               & $\sqrt{}$                      & $\sqrt{}$                   &                                & $\sqrt{}$                  & $\sqrt{}$                    & $\sqrt{}$                    & $\sqrt{}$                & $\sqrt{}$                         & $\sqrt{}$     \\ \hline
Camera image                & $\sqrt{}$                       &                         &                                &                             & $\sqrt{}$                      &                            &                              &                              &                          &                                   &               \\ \hline
Experimental evaluation     & $\sqrt{}$                       & $\sqrt{}$               &                                &                             & $\sqrt{}$                      & $\sqrt{}$                  & $\sqrt{}$                    & $\sqrt{}$                    &                          & $\sqrt{}$                         & $\sqrt{}$     \\ \hline
LOS blockage by pedestrians & $\sqrt{}$                       &                         & $\sqrt{}$                      &                             & $\sqrt{}$                      &                            &                              &                              &                          &                                   & $\sqrt{}$     \\ \hline
Look ahead prediction       & $\sqrt{}$                       & $\sqrt{}$               & $\sqrt{}$                      &                             &                                & $\sqrt{}$                  &                              &                              &                          &                                   & $\sqrt{}$     \\ \hline
\end{tabular}
\label{tab:rlwk1}
\end{table*}

\begin{table*}[t!]
\setlength\tabcolsep{3pt}
\caption{Comparison of link quality prediction methods for reliable mmWave communications}
\centering
\small
\begin{tabular}{cccccccc} \toprule
Existing works                & \cite{nishio:jsac2019proactive}                             & \cite{wu:wcnc2022lidar}                        & \cite{marasinghe:gcw2021lidar}                 & \cite{klautau:wcl2019lidar}                    & \cite{zhang:infocom2022vision}                 & \cite{asano:apcws2021high}                                  & \textbf{Ours}                                            \\ \midrule
Vision sensor            & Depth camera                                                & LiDAR                                          & LiDAR                                          & LiDAR                                          & RGB camera                                     & mmWave radar                                                & \begin{tabular}{c} Depth camera \\ \& LiDAR\end{tabular} \\ \midrule
Raw data                 & Depth image                                                 & Point cloud                                    & Point cloud                                    & Point cloud                                    & RGB image                                      & Point cloud                                                 & Point cloud                                              \\ \midrule
Feature for prediction   & Depth image                                                 & Heatmap image                                  & Bounding box                                   & 3D histogram                                   & Bounding box                                   & Link quality map                                            & Voxel grid                                               \\ \midrule
Frequency band           & 60\,GHz                                                     & 60\,GHz                                        & (mmWave/THz)                                   & 60\,GHz                                        & 28\,GHz                                        & 60\,GHz                                                     & 60\,GHz                                                  \\ \midrule
Evaluation method        & \begin{tabular}{c} Simulation \& \\ Experiment\end{tabular} & Experiment                                     & Simulation                                     & Simulation                                     & Experiment                                     & \begin{tabular}{c} Simulation \& \\ Experiment\end{tabular} & Experiment                                               \\ \midrule
Evaluation environment   & Indoor                                                      & Outdoor                                        & Indoor                                         & Outdoor                                        & Indoor                                         & Outdoor                                                     & Indoor                                                   \\ \midrule
LOS path blocker         & Pedestrian                                                  & Vehicle                                        & Pedestrian                                     & Vehicle                                        & Pedestrian                                     & Vehicle                                                     & Pedestrian                                               \\ \midrule
Prediction formulation   & Regression                                                  & Classification                                 & Classification                                 & Classification                                 & Classification                                 & Regression                                                  & Regression                                               \\ \midrule
Link quality indicator   & RSSI                                                        & \begin{tabular}{c} LOS or \\ NLOS\end{tabular} & \begin{tabular}{c} LOS or \\ NLOS\end{tabular} & \begin{tabular}{c} LOS or \\ NLOS\end{tabular} & \begin{tabular}{c} LOS or \\ NLOS\end{tabular} & RSSI                                                        & \begin{tabular}{c} RSSI \& \\ Throughput\end{tabular}    \\ \midrule
How long ahead to predict & 500\,ms                                                     & 1000\,ms                                       & 300\,ms                                        & ---                                            & ---                                            & ---                                                         & 1000\,ms                                                 \\ \bottomrule
\end{tabular}
\label{tab:rlwk2}
\end{table*}

Our previous study~\cite{nishio:jsac2019proactive} used camera images as the key enabler of proactive mmWave link quality prediction.
Camera images capture vision information about the environment and thus contain information necessary to predict mmWave communication LOS path blockage.
The mmWave communication environment was captured by a depth camera, and ML was used to predict future received signal strength indicator (RSSI) using time series data of depth images.
Three ML algorithms were used: two NN models, including convolution layers and convolutional LSTM layers~\cite{shi:nips2015convlstm}, and random forest~\cite{leo:2001randomforest}.
Experiments were conducted indoors in a scenario in which a 60\,GHz band mmWave communication LOS path was blocked by pedestrians.
Experimental evaluation results show that large attenuations of the RSSI 500\,ms ahead can be predicted.
This result suggests that ML models can learn the relationship between the movement information of obstacles to the LOS path in the time series of camera images and future link quality.
However, as mentioned in \cref{s:intro}, camera images may contain private information, such as human faces, text from documents, or computer screens. 
As a result, the image-based RSSI prediction system~\cite{nishio:jsac2019proactive} may be difficult to implement in locations with strict privacy-related constraints, such as private homes, company offices, and hospitals. 
To address this issue, non-image-based link quality prediction methods can be employed. 
Our proposed solution utilizes point clouds, which are sparser than images and significantly reduce the likelihood of identifying personal or sensitive information.

Wu et al.~\cite{wu:wcnc2022lidar} proposed a mmWave blockage prediction method using ML and point cloud. 
In this work, point clouds, which observe a mmWave communication area, are converted into heatmap images by calculating the distance to the reflection point for each horizontal angle and arranging them in the time direction. 
From these heatmaps, a binary link state, either LOS or NLOS, is predicted.
An experiment was conducted in a scenario in which the mmWave communication LOS path is blocked by vehicles in an outdoor environment.
The NN model includes convolution, which learns the relationship between the heatmap image and the binary label.
Based on the evaluation results, the system predicts blockages that occur within 100\,ms with 95\% accuracy and blockages that occur within 1000\,ms with more than 80\% accuracy.

Marasinghe et al.~\cite{marasinghe:gcw2021lidar} proposed a mmWave communication LOS path blockage prediction method by detecting human position and motion from point clouds and predicting future positions.
An NN model, including LSTM layers~\cite{hochreiter1997lstm} was used to predict the future human bounding box, and a ray tracing algorithm was used to predict a binary link state, either LOS or NLOS.
This method was evaluated through computer simulations in a scenario in which the mmWave or terahertz (THz) communication LOS path was blocked by humans in an indoor environment.
The system could predict future blockages with an accuracy of 87\% while maintaining 78\% precision and 79\% recall 300\,ms ahead.

Klautau et al.~\cite{klautau:wcl2019lidar} proposed a point cloud-based LOS blockage prediction method for mmWave beam selection.
In this method, LiDAR point clouds observed in a wireless communication environment were converted into 3D histograms, and LOS probability was inferred using a convolutional NN.
This method has been evaluated to discriminate LOS with a 90\% accuracy through simulations in a vehicle-to-infrastructure (V2I) scenario.

Zhang et al.~\cite{zhang:infocom2022vision} demonstrated a platform for beam tracking and blockage prediction, using stereo cameras and LiDAR for mmWave communications and frequency switching from mmWave to sub-6\,GHz just before a blockage occurs.
In particular, objects blocking the mmWave LOS path were detected based on RGB images, and the blockage was predicted using a NN with recurrent layers from the time series of bounding boxes.
The transmitter (Tx) and receiver (Rx) were simultaneously detected from the LiDAR point cloud and used for beam tracking.
These methods~\cite{wu:wcnc2022lidar, marasinghe:gcw2021lidar, klautau:wcl2019lidar, zhang:infocom2022vision} focus on predicting the binary link state, either LOS or NLOS, whereas our proposed method is capable of quantitatively predicting link quality in environments where the link quality dynamically changes due to human blockage.

Asano et al.~\cite{asano:apcws2021high} proposed a transmission timing control method with link quality prediction for mmWave vehicular to infrastructure (V2I) communications. 
In this method, the link quality is predicted using simulations from the positions of the vehicles obtained by mmWave radars and the given location of roadside units. 
The transmission timing control method can improve the average throughput in mmWave V2I communications. 
This study primarily discusses timing control methods and vehicle position prediction, without sufficiently addressing link quality prediction. 
Furthermore, it mainly focuses on V2I communications, which is a significantly different environment from the indoor communications we are targeting.

We also summarize some studies not included in \cref{tab:rlwk2} but mentioned in \cref{tab:rlwk1}.
Egi et al.~\cite{egi:access2019machine} proposed an ML-based path loss estimation method for the 1.8\,GHz band in outdoor environments. 
This method leverages satellite images and point clouds to estimate path loss attenuation caused by static obstacles such as trees and buildings.
Järveläinen et al.~\cite{jan:lwc2016evaluation} proposed a LOS probability prediction method for outdoor communications with accurate point clouds of the outdoor environment. 
Järveläinen et al.~\cite{jan:tap2016indoor} also proposed a point cloud-based ray-tracing simulation method to estimate the indoor propagation of mmWave channels.
Stéphan et al.~\cite{stephan:eucap2018increased} proposed utilizing accurate geographical data such as point clouds acquired by LiDAR to increase the reliability of mmWave link simulations in outdoor environments.

As summarized above, several existing studies have investigated link quality prediction and the use of point clouds for wireless communications. 
However, these studies do not focus on the quantitative and deterministic prediction of the future link quality of mmWave communications, which dynamically changes due to human blockage, from point clouds instead of camera images.
\section{POINT CLOUD-BASED LINK QUALITY PREDICTION}\label{s:proposedmethod}
\subsection{SYSTEM MODEL}\label{ss:system_model}
\begin{figure}[t!]
    \centering
    \includegraphics[width=85mm]{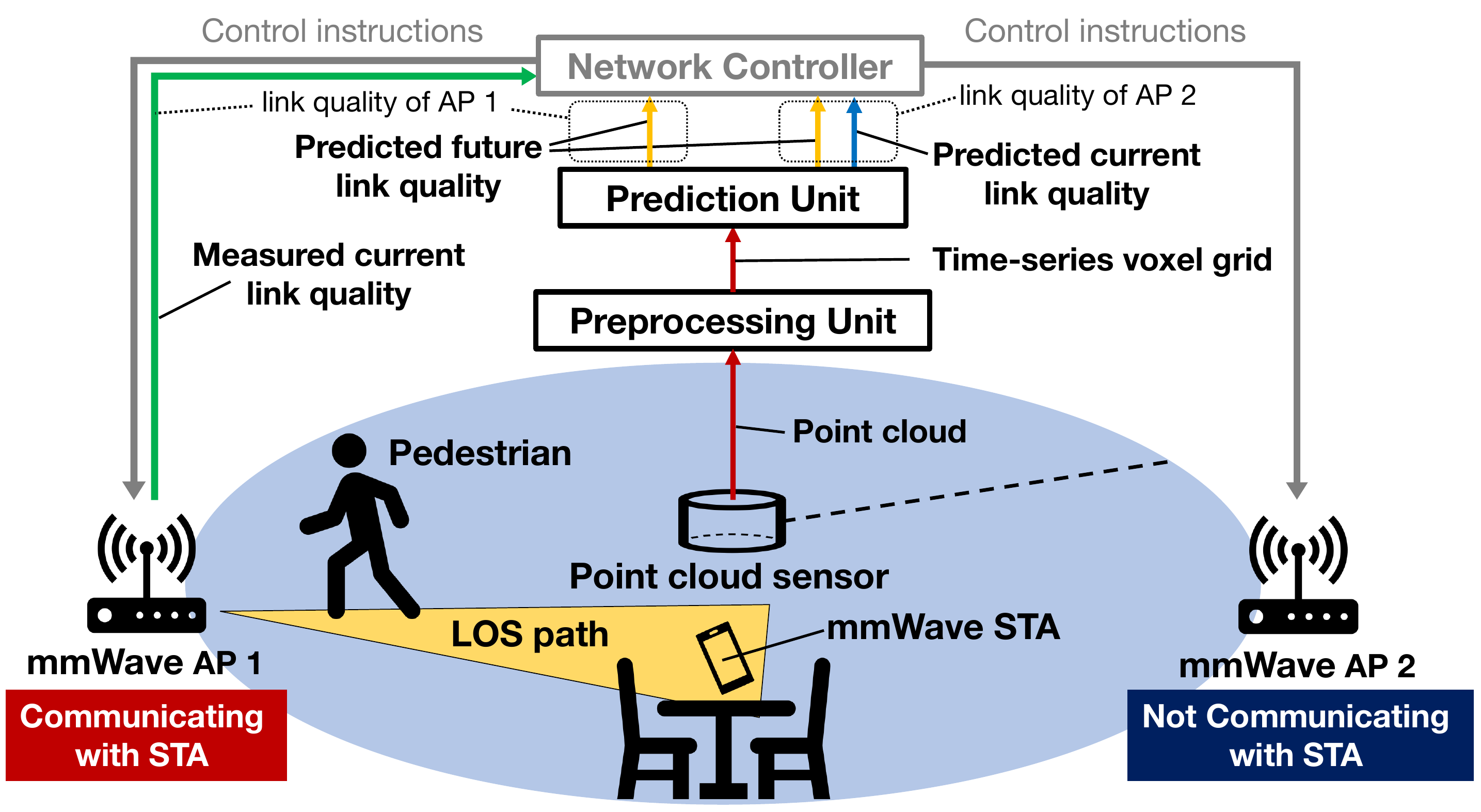}
    \caption{System model of point cloud sensor-aided mmWave communication system, which aims to proactively control wireless links by predicting link quality using point clouds.}
    \label{fig:system_model}
\end{figure}

\cref{fig:system_model} illustrates the system model of the point cloud sensor-aided mmWave communication system, which aims to proactively control wireless links based on the proposed link quality prediction using point clouds.
This system consists of APs and STAs for mmWave communication, a point cloud sensor, such as LiDAR or depth camera or radar, to observe the environment and obtain point clouds, a preprocessing unit, a prediction unit, and a network controller.
On the STA, applications requiring large amounts of data are in use, and the AP and STA generate large amounts of traffic through mmWave communications.
The point cloud sensor acquires and transmits point clouds to the preprocessing unit.
The preprocessing unit reduces data volume and noise of point clouds, and converts data format.
The details of the preprocessing unit are described in Section~III-D.
The prediction unit infers future and current link quality values, such as RSSI or throughput, using ML.
The details of the prediction unit are described in Section~III-E.
The AP measures and reports the link quality value to the network controller. 
The network controller instructs APs to take appropriate communication control actions, such as handover, beamforming, frequency switching, and IRS control based on the predicted and current link quality before the link quality deteriorates significantly due to LOS blockage.
The above-mentioned proactive communication control enables reliable mmWave communications.
The system model is consistent with the system that replaced the camera with the point cloud sensor in our previous study~\cite{nishio:jsac2019proactive}.

We assume an indoor scenario such as residences, offices, and public facilities. 
Pedestrians in the mmWave radio propagation space move aperiodically, and the mobility is observed as point clouds obtained by the point cloud sensor.
In mmWave communications, link quality (i.e., RSSI and throughput) is significantly degraded when the LOS path is blocked by pedestrians.
This study assumes a simple case in which the AP and STA do not move, and the LOS between the AP and STA is within the field of view (FOV) of the point cloud sensor.
Therefore, the point cloud is expected to contain the essential visual information of mmWave radio propagation in mmWave communications.

\begin{figure}[t!]
    \centering
    \includegraphics[width=85mm]{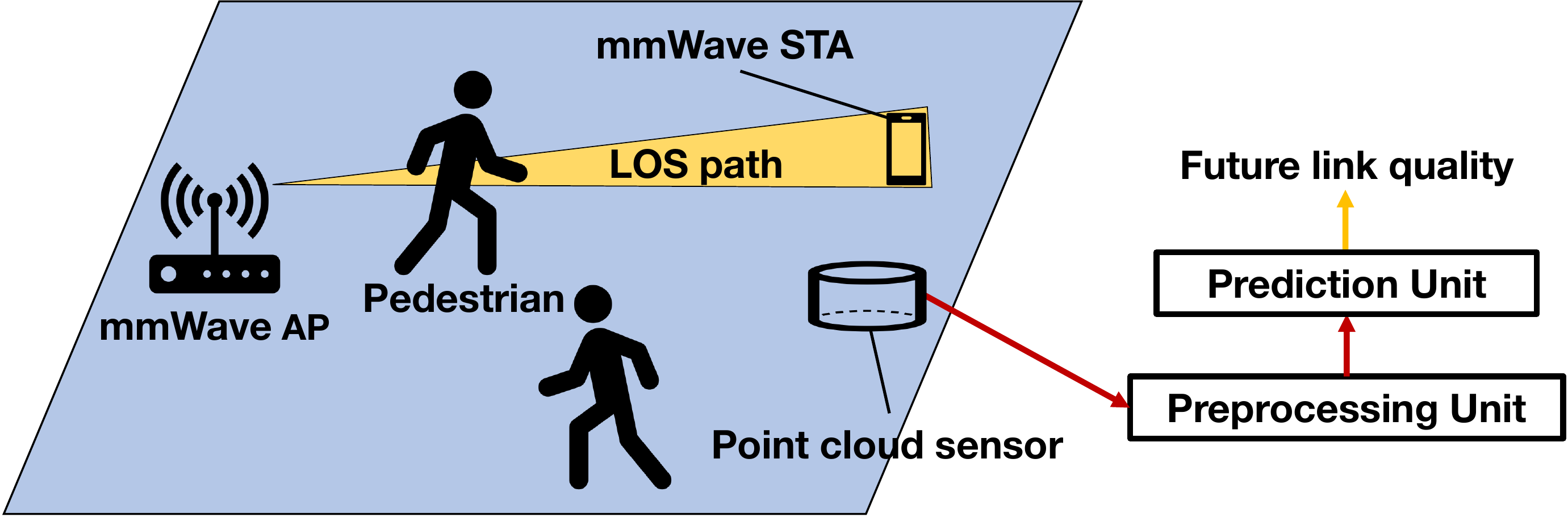}
    \caption{Simplified system model that focuses on link quality prediction. Predicting the link quality between a single AP and STA in situations where the link quality dynamically changes due to human blockage.}
    \label{fig:system_model_simple}
\end{figure}

The objective of this paper is to demonstrate the feasibility of mmWave link quality prediction using point clouds.
Therefore, we assume a simplified system model that focuses only on link quality prediction hereinafter, as shown in \cref{fig:system_model_simple}.
We aim to predict link quality between one AP and one STA based on point clouds in situations where the link quality dynamically changes due to human blockage.

\subsection{FORMULATION OF POINT CLOUD-BASED LINK QUALITY PREDICTION}\label{ss:formulation}
This study aims to quantitatively and deterministically predict future link quality (i.e., RSSI or throughput) from time series data of point clouds representing radio propagation spaces of mmWave communications.
This problem can be formulated as a regression that maps from the time series data of a point cloud to future link quality values.
Let $n$ be the number of points included in a point cloud and $\bm{p_i} = (x_i, y_i, z_i)$ be the coordinates of the $i$-th point, the point cloud $\mathcal{P}$ is as follows:
\begin{align}
    \mathcal{P} = \bigcup_{i=0}^{n-1} \left\{(x_i, y_i, z_i)\right\}.
\end{align}
Note that the points in the point cloud are in no particular order.
Let $\mathcal{P}_t$ be the point cloud at timestep $t$, the time series data $\mathcal{D}_{t,s}$ of the point cloud for the previous $s$ timesteps at some timestep $t$ is as follows:
\begin{align}
    \mathcal{D}_{t,s} = \left( \mathcal{P}_{t-s+1}, \mathcal{P}_{t-s+2}, \cdots, \mathcal{P}_{t}\right).
\end{align}
Let $q_t \in \mathbb{R}$ be the link quality value at a particular timestep $t$, $q_{t+k}$ represents $k$ timesteps ahead link quality value from a particular timestep $t$.
The regression task can be formulated as the problem of determining the parameterized mapping function $f_W$, which maps from a time series of point clouds $\mathcal{D}_{t,s}$ to the link quality $k$ timesteps ahead, $q_{t+k}$. In other words, the problem is to find a function satisfying
\begin{align}
    q_{t+k} = f_W(\mathcal{D}_{t,s})\label{eq:formulation}.
\end{align}

The mapping function $f_W$ is obtained by solving a minimization problem that seeks to minimize a loss function $l$, which measures the magnitude of the error between the true value $q_{t+k}$ and the predicted value $f_W(\mathcal{D}_{t, s})$, with respect to the parameter $W$. The problem can be expressed as follows:
\begin{mini}
{\substack{W}}
{\sum_{t \in \mathcal{T}} l\left(q_{t+k}, f_W(\mathcal{D}_{t, s})\right)\label{eq:loss}.}
{}{}
\end{mini}
Here, $\mathcal{T}$ represents the set of timesteps $t$ included in a training set $\mathcal{S}_\mathrm{train}$ (i.e., the timesteps where both $q_{t+k}$ and $\mathcal{D}_{t, s}$ are available).
%We call the process of solving loss minimization problem~\cref{eq:loss} \texttt{training process}.
In this paper, we used $L^2$ losses such as mean-squared error (MSE) and RMSE as the loss function. 
In predicting mmWave link quality, it is important to predict significant variations caused by human blockages, rather than minor fluctuations during LOS communication. 
The property of $L^2$ loss, which emphasizes larger errors compared to $L^1$ loss, is expected to mitigate the large errors that arise when predicting high link quality during blockages or low link quality during LOS communication.

\subsection{PREDICTION METHOD DESIGN POLICY}\label{ss:policy}
We use a supervised learning framework to solve the aforementioned regression task.
Generally, supervised learning requires a large number of labeled datasets to obtain an accurate mapping. 
In the proposed system, labeling can be automatic, using the observed link quality, and constructing a large labeled dataset is easy, unlike tasks that require manual labeling, such as object detection and segmentation.

\begin{figure}[t!]
    \centering
    \includegraphics[width=85mm]{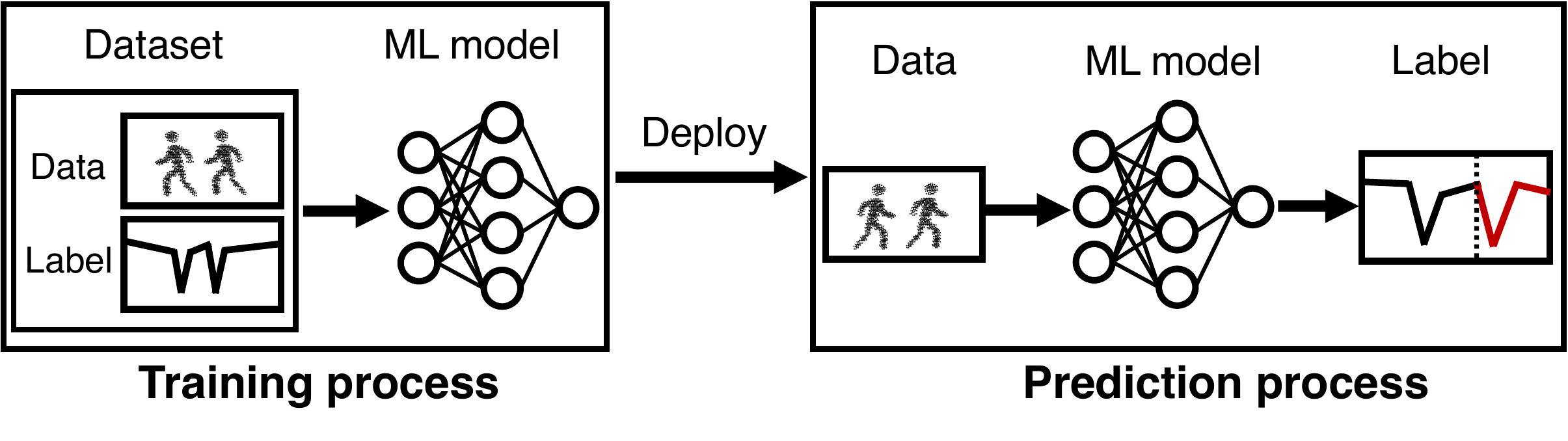}
    \caption{\texttt{Training process} and \texttt{prediction process}. Point cloud and RSSI/throughput are used for data and labels.}
    \label{fig:process}
\end{figure}

The proposed system consists of two main processes: the \texttt{training process} and the \texttt{prediction process}, as illustrated in \cref{fig:process}. 
In the \texttt{training process}, a supervised learning model is trained using a labeled dataset to learn the correspondence between the input data (i.e., point cloud) and labels (i.e., RSSI and throughput) by solving the aforementioned minimization problem \eqref{eq:loss}.
The \texttt{training process} is initiated when the training dataset is available, which involves the transmission of signals by the STA through the mmWave link, obtaining the measured link quality from the AP, and capturing the point cloud from the sensor. 
The details of the labeled dataset creation method are presented in Section~III-D. 
Once the model is trained, the \texttt{prediction process} can be carried out anytime, given the availability of time series of point clouds, which predicts RSSI and throughput from the point cloud data.
Notably, the model can be updated whenever data is accumulated, even after its deployment. However, for the sake of simplicity, we did not consider retraining the model in this paper and left it as future work.

Varieties of supervised learning algorithms exist, and we selected the appropriate algorithm for link quality prediction.
A simple approach is to apply deep learning models specialized in point clouds, such as PointNet~\cite{qi:cvpr2017pointnet, qi2017pointnet++} and VoteNet~\cite{qi2019vote}, which can directly input point clouds.
These point cloud-based models can directly map from point clouds to the target values.
However, based on our preliminary experiments, these models could not predict link quality.
Therefore, this study adopted a method that is a 3D extension of the method used in the previous study~\cite{nishio:jsac2019proactive}.
Specifically, point clouds are converted into voxel grid data format that can be handled by convolution layers and convolutional LSTM layers.

The proposed method consists of preprocessing that converts point clouds to voxel grid format and an ML model to learn a mapping from the time series voxel grid to the future link quality value such as RSSI and throughput. 
Section~III-D and Section~III-E describe the preprocessing method and ML model, respectively.

\subsection{PREPROCESSING METHODS FOR POINT CLOUDS}\label{ss:preprocessing}
The preprocessing unit applies downsampling and denoising to the raw point clouds acquired by LiDAR or depth cameras, which tend to be large in size and contain noise. 
Additionally, the preprocessing unit converts the point cloud data into voxel grid format to enable the application of the convolution-based algorithm. 
The proposed preprocessing method comprises six phases, including cuboid cropping, random downsampling, statistical outlier removal, voxelization, time series concatenation, and labeling.

The first three processes are used to reduce the data volume and noise.
Cropping removes redundant regions of the point cloud.
Point clouds obtained from LiDAR sometimes cause inaccurate point observations due to the effects of reflective objects such as windows and mirrors.
Limiting the region based on prior geographic knowledge can remove such obvious noise and points in regions that are irrelevant to sensing. 
We employed the cuboid cropping method to cut out a rectangular region from a 3D point cloud. 
Specifically, for each $i$-th point $\bm{p_i} = (x_i,y_i,z_i)$ in a point cloud, the point is removed when the $(x_i,y_i,z_i)$ coordinates are outside the rectangle region $[x_{\min}, y_{\min}, z_{\min}] \times [x_{\max}, y_{\max}, z_{\max}]$.

Random downsampling reduces the number of points by randomly selecting a subset of points from the original point cloud according to a reduction rate $r_\mathrm{d}$.
Specifically, it arranges the points in random order and only retains the points whose indices are up to the product of the original number of points and the reduction rate $r_\mathrm{d}$. 
However, there is a tradeoff between reducing the number of points to reduce computational complexity and preserving spatial features. 
A low reduction rate may reduce computational complexity but remove important spatial features.
In this study, we considered the reduction rate $r_\mathrm{d}$ as a hyperparameter and experimentally determined its optimal value in \cref{s:rssi} and \cref{s:thrp}. 
However, we did not explore other methods for finding the optimal reduction rate as it was beyond the scope of this study.

Outlier removal removes noise points resulting from the measurements.
Outlier removal enables an accurate understanding of the 3D space.
Statistical outlier removal is a method of removing points that are far from their neighbors by comparing the average distance between all points.
Statistical outlier removal is employed in this method because it can remove outliers independent of the scale of the region in which points exist.
Two hyperparameters exist for statistical outlier removal: the number of nearest neighbor points $n_\mathrm{o}$ and the standard deviation ratio $r_\mathrm{o}$.
First, the mean $\mu$ and standard deviation $\sigma$ of the distances between all points are calculated.
Subsequently, for each $i$-th point $\bm{p_i}$, the average distance $d_i$ to $n_\mathrm{o}$-nearest neighbor points is calculated.
Finally, if $\mu + r_\mathrm{o}\sigma < d_i$, the point $\bm{p_i}$ is removed.
These two hyperparameters $n_\mathrm{o}$ and $r_\mathrm{o}$ were experimentally determined in this study. 

Voxelization divides the space where point clouds exist by voxels, which are 3D extensions of pixels, and arranges them into a voxel grid, which is the regular grid in 3D space.
A voxel grid can be represented as a 3D array on the computer, and the voxel grid can be input to the ML model proposed in Section~III-E.
The detailed voxelization method is described in Appendix~A.
The shape of the voxel grid is calculated using the voxel size $s_\mathrm{v}$ and observation environment.
To convert a point cloud into a voxel grid while preserving spatial characteristics, the voxel size $s_\mathrm{v}$ must be appropriately determined while considering the observation space.
In this study, the voxel size $s_\mathrm{v}$ is treated as a hyperparameter and was experimentally determined owing to the capability of localizing pedestrians in the experimental environment.
Open3D~\cite{zhou:arxiv2018open3d} and Point Cloud Library~\cite{rusu:icra2011pcl} are used for cuboid cropping, random downsampling, statistical outlier removal, and voxelization.

Subsequently, the voxel grids are concatenated in the time direction to generate time series data.
As formulated in Section~III-C, the previous $s$ timestep data is concatenated.
Specifically, a time series data $\mathcal{D}_{t,s}$ is generated by concatenating 3D arrays representing voxel grid data generated in the previous preprocessing steps.
After the concatenation, the time series data $\mathcal{D}_{t,s}$ becomes a 4-dimensional array with the shape of $(s, h, w, d)$.
The parameter $s$ represents the number of past frames concatenated for the time series input, and $h$, $w$, and $d$ represent the height, width, and depth of the voxel grid, respectively.

Finally, we describe the generation of labeled datasets for the \texttt{training process} introduced in Section~III-C.
We used temporal difference labeling proposed in \cite{nishio:jsac2019proactive} for data annotation.
Specifically, for all timesteps $t$, we map the voxel grid time series data $\mathcal{D}_{t,s}$ to the $k$ timesteps ahead link quality value $q_{t+k}$; thus, a labeled sample is generated as pair like $(\mathcal{D}_{t,s}, q_{t+k})$.
The use of labeled datasets enables the training of an ML model that predicts $k$ timesteps ahead of future link quality value in Section~III-E.

\cref{tab:example_prepro} shows an example of preprocessing results for the actual experimental data.
In this paper, two types of point clouds are used for the experimental evaluation: depth camera point cloud and LiDAR point cloud.
Both point clouds are observations of two people blocking LOS paths of mmWave communication in an indoor environment.
Depth camera point clouds tend to have a lot of noise in the raw data.
LiDAR point clouds tend to have noise points at locations that are far outliers.
These noises disappeared by applying cuboid cropping, random downsampling, and statistical outlier removal.
After voxelization, both point clouds were converted into voxel grids while preserving the shape of the human bodies.

\begin{table}[t!]
  \caption{Examples of preprocessed point clouds and their bounding boxes}
  \centering
  \begin{tabular}{cll} \toprule
    Point cloud type & Depth camera point cloud & LiDAR point cloud \\ \midrule
    Raw point cloud &
    \begin{minipage}{0.35\columnwidth}
      \centering
      \scalebox{0.125}{\includegraphics{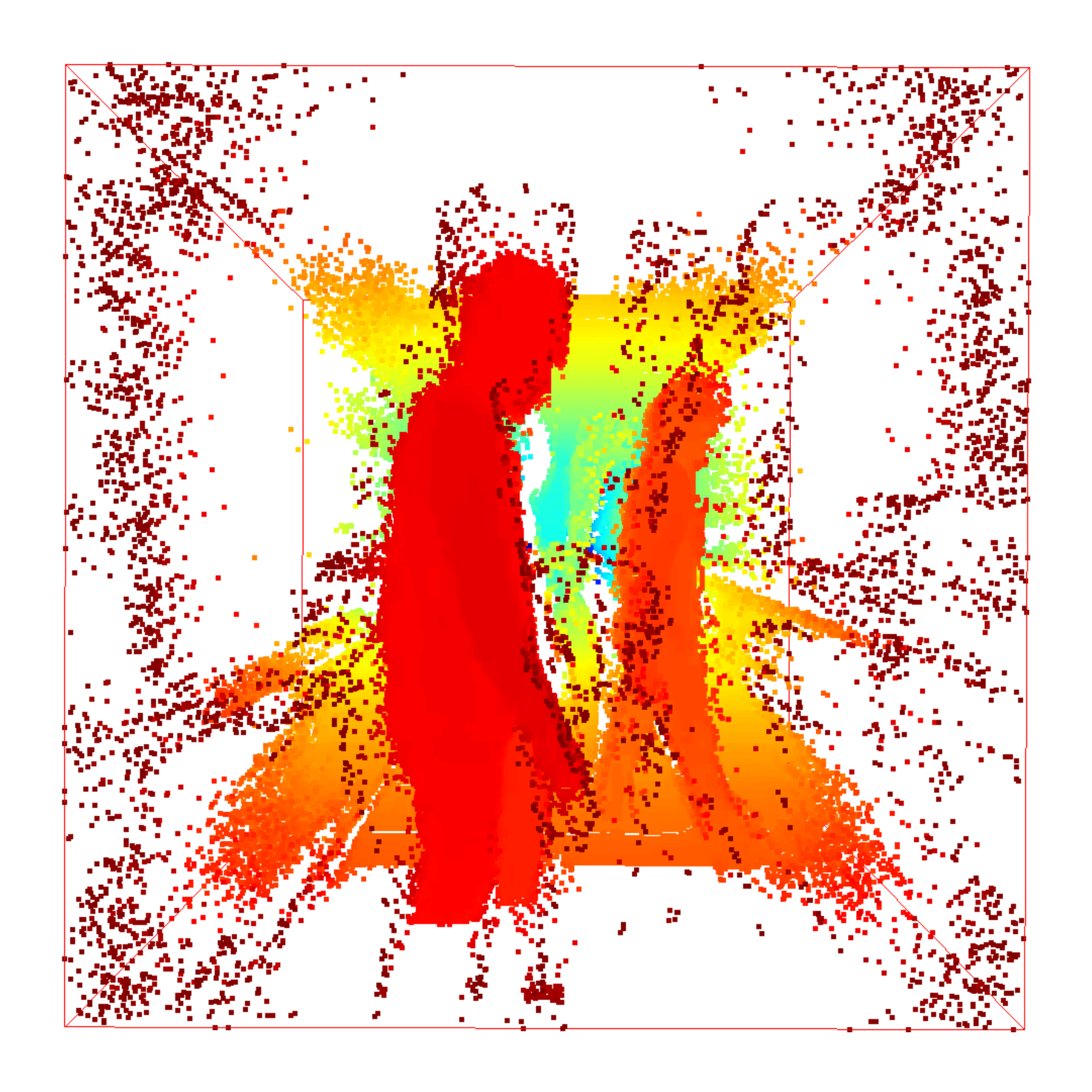}}
    \end{minipage} &
    \begin{minipage}{0.1\columnwidth}
      \centering
      \scalebox{0.125}{\includegraphics{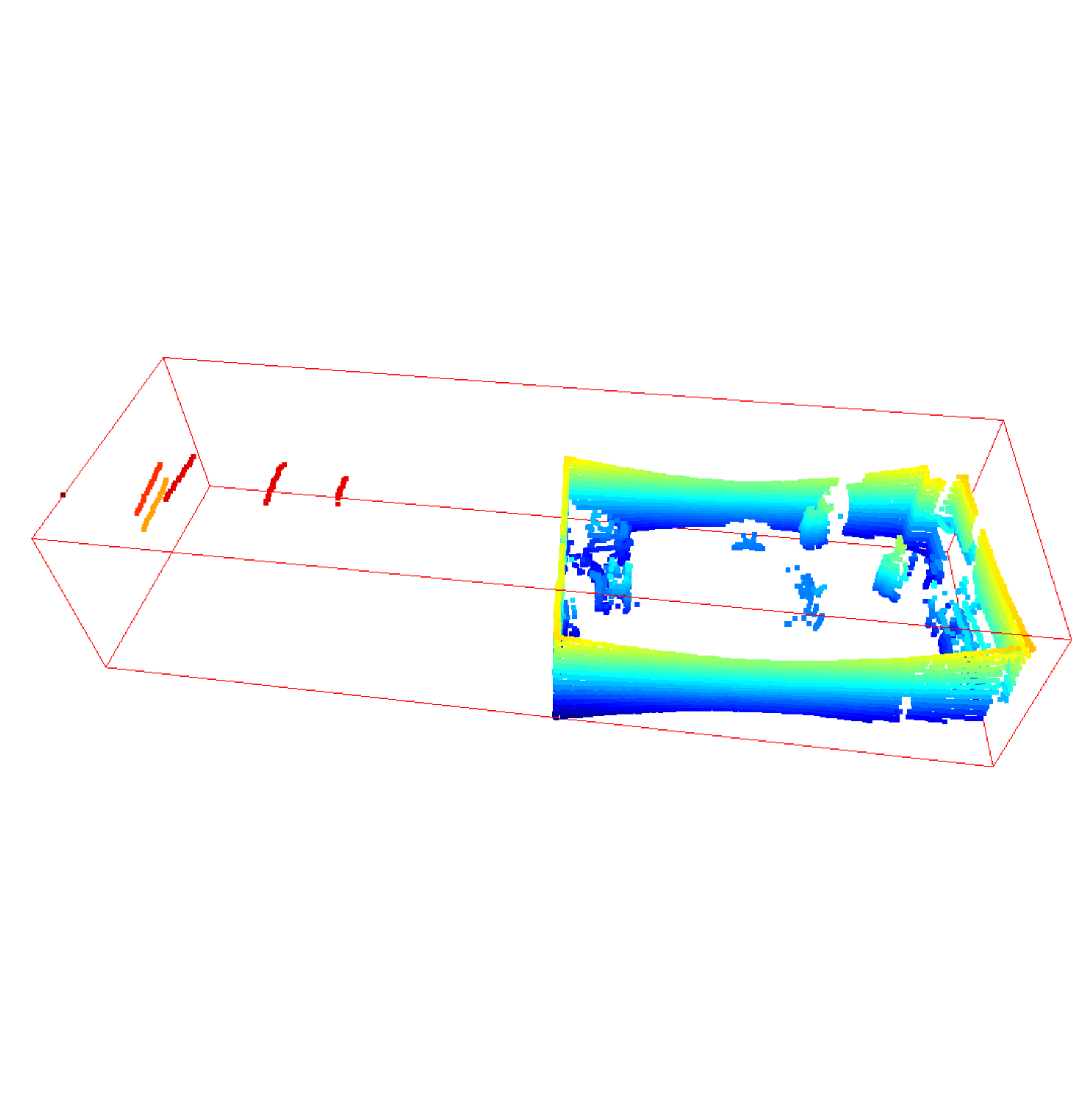}}
    \end{minipage} \\ \midrule
    \begin{minipage}{20truemm} 
    After\\cuboid cropping,\\random downsampling,\\statistical\\outlier removal
    \end{minipage}&
    \begin{minipage}{0.35\columnwidth}
      \centering
      \scalebox{0.125}{\includegraphics{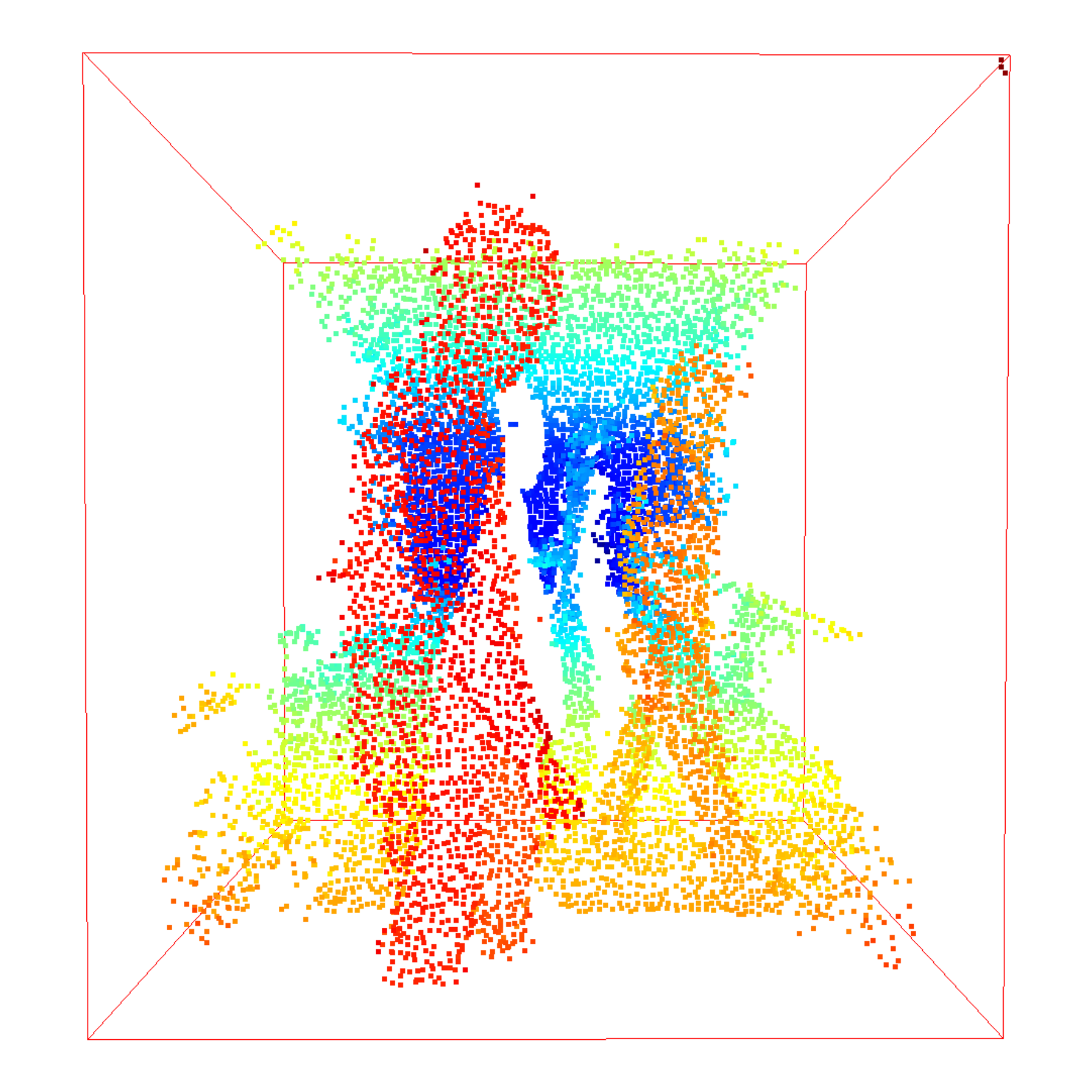}}
    \end{minipage} &
    \begin{minipage}{0.1\columnwidth}
      \centering
      \scalebox{0.125}{\includegraphics{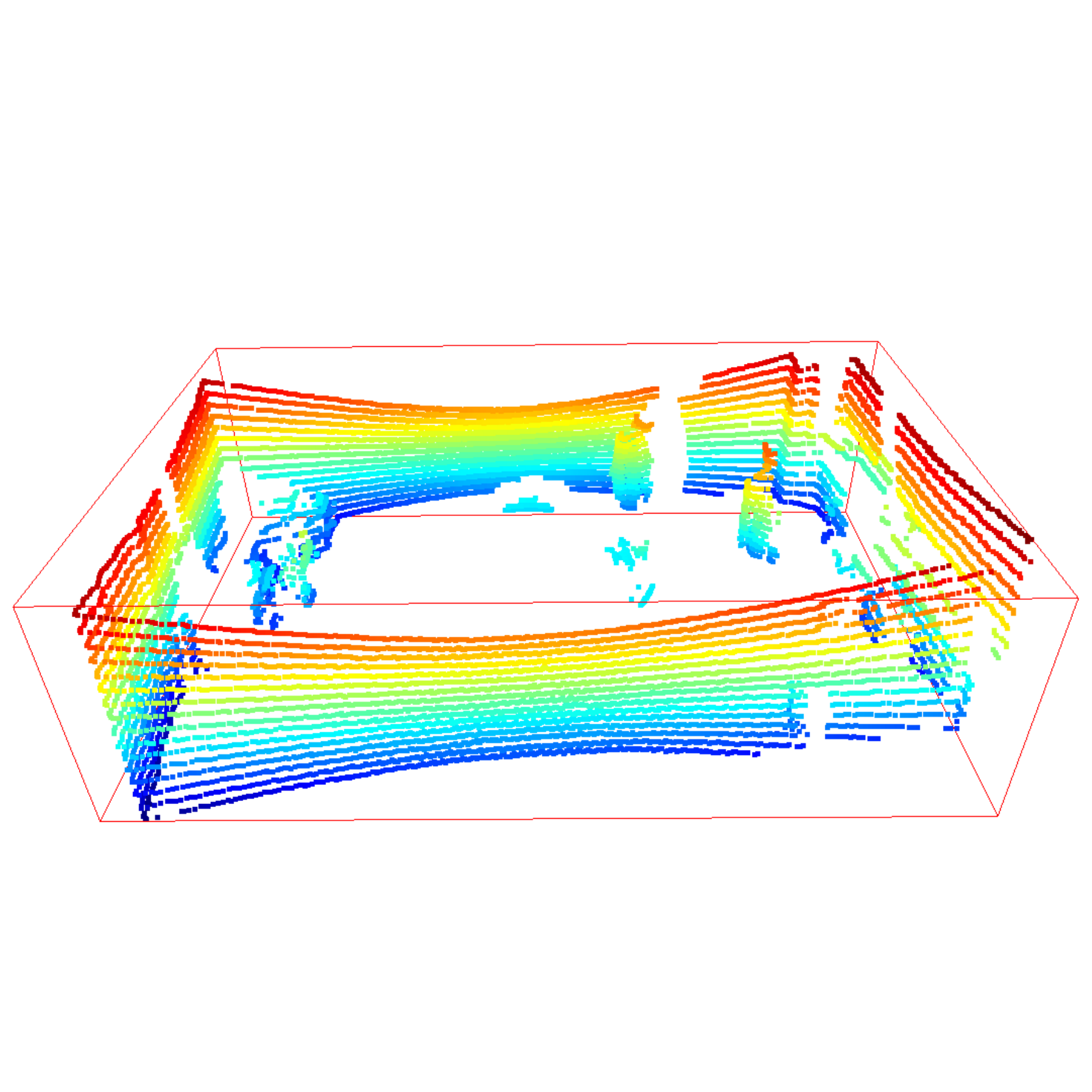}}
    \end{minipage} \\ \midrule
    After voxelization &
    \begin{minipage}{0.35\columnwidth}
      \centering
      \scalebox{0.125}{\includegraphics{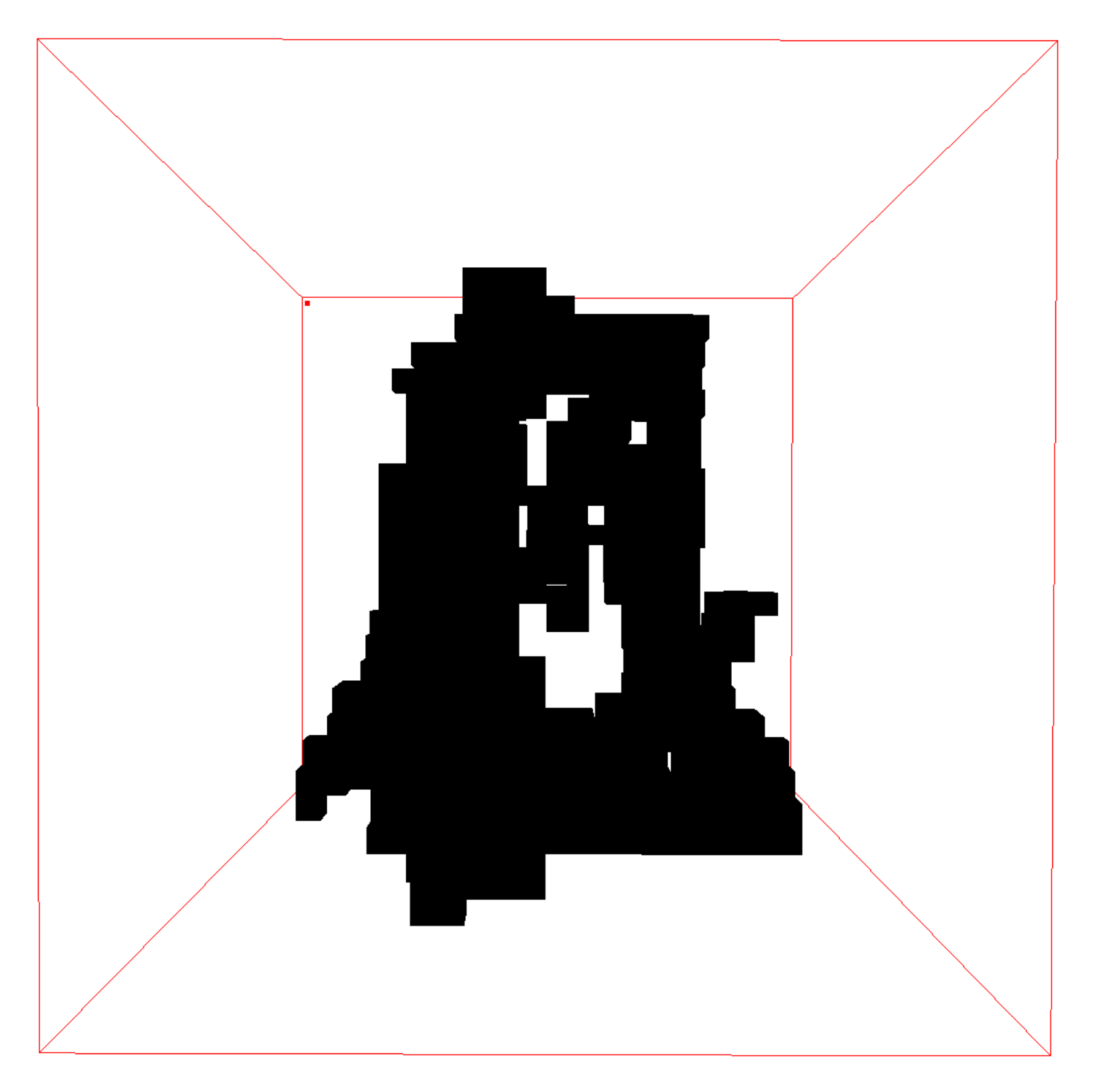}}
    \end{minipage} &
    \begin{minipage}{0.1\columnwidth}
      \centering
      \scalebox{0.125}{\includegraphics{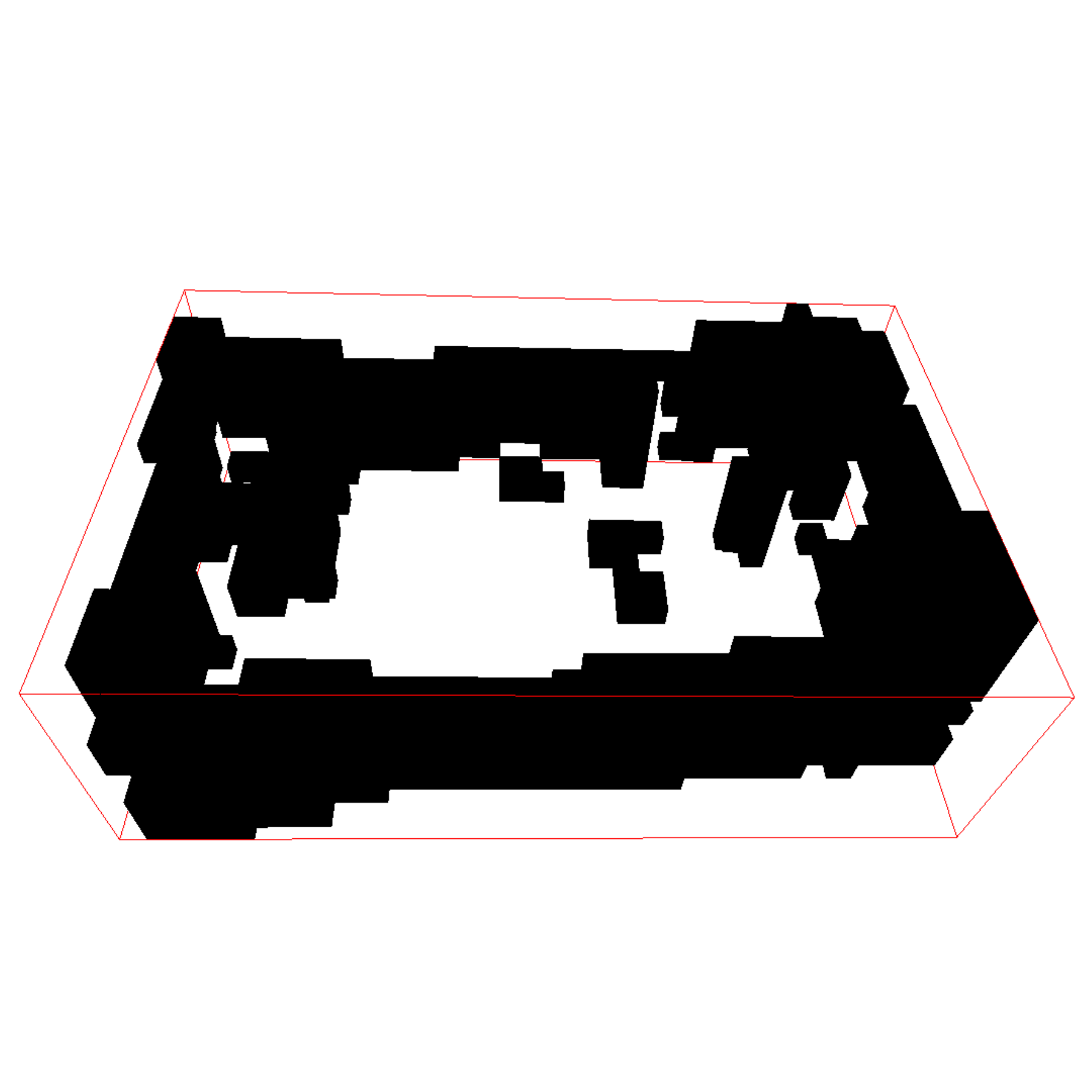}}
    \end{minipage} \\ \bottomrule
  \end{tabular}
  \label{tab:example_prepro}
\end{table}

\subsection{MACHINE LEARNING METHODS}\label{ss:ml}
The voxel grid generated during the preprocessing process serves as the input for the ML model that maps the voxel grid to link quality values. 
The voxel grids contain the spatial information of objects, and by using a time series voxel grid, temporal information can also be obtained. 
The spatio-temporal information, including the location and motion of objects, is crucial for accurate link quality prediction.
Many ML models have already been proposed for the computer vision task of extracting spatio-temporal features from voxel grids.
We examined three ML models: a NN with 3D convolution layers (Conv3D), a NN with 3D convolutional LSTM layers (ConvLSTM3D), and GBDT.
Although ML algorithms that outperform the algorithms used in this study may exist, a comprehensive investigation of the ML algorithms and hyperparameter tunings is beyond the scope of this study, because this study aims to design a mechanism to predict the mmWave link quality value and to demonstrate its feasibility.

\begin{figure}[t!]
  \begin{minipage}[b]{0.49\columnwidth}
    \centering
    \includegraphics[width=\columnwidth]{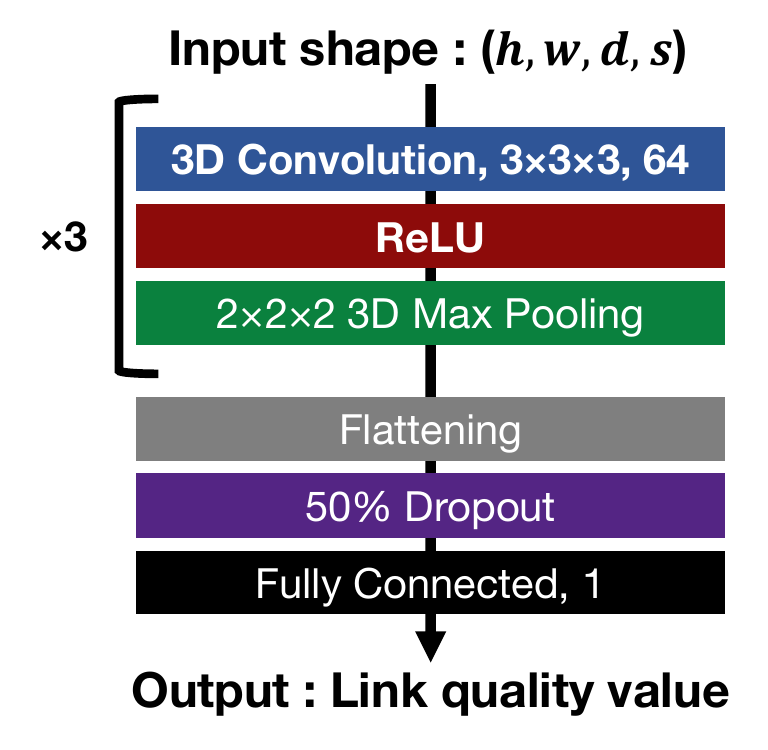}
    \subcaption{Conv3D}
  \end{minipage}
  \begin{minipage}[b]{0.49\columnwidth}
    \centering
    \includegraphics[width=\columnwidth]{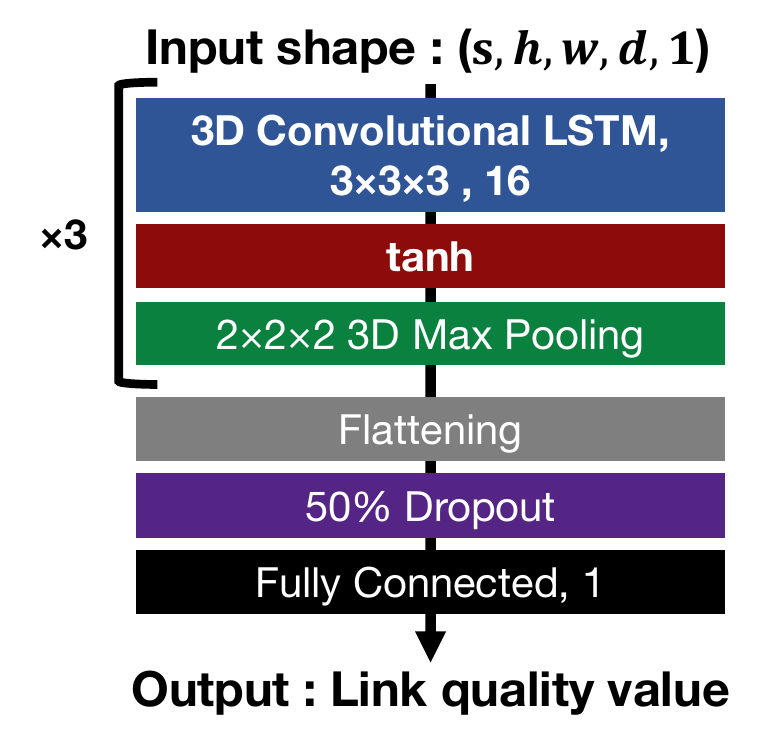}
    \subcaption{ConvLSTM3D}
  \end{minipage}
  \caption{Two NN structures. 
    Parameters $h$, $w$, and $d$ are the height, width, and depth of the voxel grid, respectively, and $s$ is the number of past frames concatenated for the time series input.
    Conv3D and ConvLSTM3D differ in that the input shape is $(h,w,d,s)$ and $(s,h,w,d,1)$, the first layer is 3D convolution and 3D convolutional LSTM, the activation function is ReLU and $\tanh$, respectively.}
    \label{fig:nn_structure}
\end{figure}

\cref{fig:nn_structure} depicts the NN architectures used in this study, named Conv3D and ConvLSTM3D, which are 3D extensions of our previously proposed models for image-based prediction~\cite{nishio:jsac2019proactive}, named CNN and CNN+ConvLSTM, respectively.
Conv3D and ConvLSTM3D can input 3D data (i.e., voxel grids) and are both designed to extract spatio-temporal features but in different ways.

The Conv3D model consists of the following layers: 3D convolution, rectified linear unit (ReLU), 3D max pooling, flattening, dropout~\cite{srivastava:jmlr2014dropout}, and fully connected layer.
We designed Conv3D model layer architectures based on the dense voxel grid-based method proposed for object recognition and robot control~\cite{maturana:iros2015voxnet}.
The input tensor shape of Conv3D is $(h, w, d, s)$, where $h$, $w$, and $d$ are the height, width, and depth of the voxel grid, respectively, and $s$ is the number of past frames concatenated for the time series input.
First, 3D convolution is applied to the voxel grid to extract spatio-temporal features.
Although 3D convolution is usually used to extract only spatial features from one voxel grid~\cite{maturana:iros2015voxnet}, our Conv3D model architecture extracts spatio-temporal features by inputting time series into the voxel channels.
ReLU and max pooling are layers used as a set with the convolution layer.
After three convolution iterations, the four-dimensional tensor is flattened to one dimension.
After flattening, a dropout layer is inserted to prevent overfitting and improve robustness.

The ConvLSTM3D model consists of the following layers: 3D convolutional LSTM~\cite{shi:nips2015convlstm}, $\tanh$, 3D max pooing, flattening, dropout, and fully connected layer. 
Hence, the differences between the ConvLSTM3D and Conv3D models are that the input tensor shape is changed from $(h, w, d, s)$ to $(s, h, w, d, 1)$, the 3D convolution layers are replaced by 3D convolutional LSTM layers, and their activation function is switched from ReLU to $\tanh$.
The ConvLSTM3D input tensor shape can be adapted by simply reshaping of Conv3D input tensor.
The convolutional LSTM layer combines the advantages of convolutional and recurrent layers, i.e., it can extract spatio-temporal features simultaneously. 
Therefore, the 3D convolutional LSTM layer can be used to process 3D data with temporal dependencies, such as time-series voxel grids and video of 3D medical images.
In particular, the use of LSTM tends to extract long-term features, but the 3D convolutional LSTM is computationally more expensive than the use of 3D convolutional layer and may require a large amount of training data to achieve high performance~\cite{shi:nips2015convlstm}.

GBDT is an ensemble learning algorithm that has achieved remarkable results in various ML tasks~\cite{ke:nips2017lightgbm}.
Similar to the existing work~\cite{nishio:jsac2019proactive} that used random forests~\cite{leo:2001randomforest}, we flatten the time series voxel grids into one-dimensional arrays and input them into the GBDT model since GBDT does not support multidimensional tensor inputs.

\section{EXPERIMENTAL EVALUATION USING DEPTH CAMERA POINT CLOUD AND MILLIMETER-WAVE RECEIVED SIGNAL STRENGTH DATASET}\label{s:rssi}
\subsection{DATASETS}\label{ss:rssi_dataset}
We evaluated our point cloud-based link quality prediction method using a depth camera point cloud dataset labeled with RSSI values of IEEE 802.11ad mmWave communications. 
The depth camera point cloud dataset was generated from the depth image dataset originally created in \cite{nishio:jsac2019proactive} by converting depth images to point clouds using the method detailed in Appendix~B.
Depth camera point clouds represent spatial information in a specific direction, similar point clouds also can be obtained from solid-state LiDAR~\cite{raj2020lidar}.
We compared the prediction result of our proposed point cloud-based method with that of the depth image-based method~\cite{nishio:jsac2019proactive}.

\begin{figure}[t!]
    \centering
    \includegraphics[width=8.5cm]{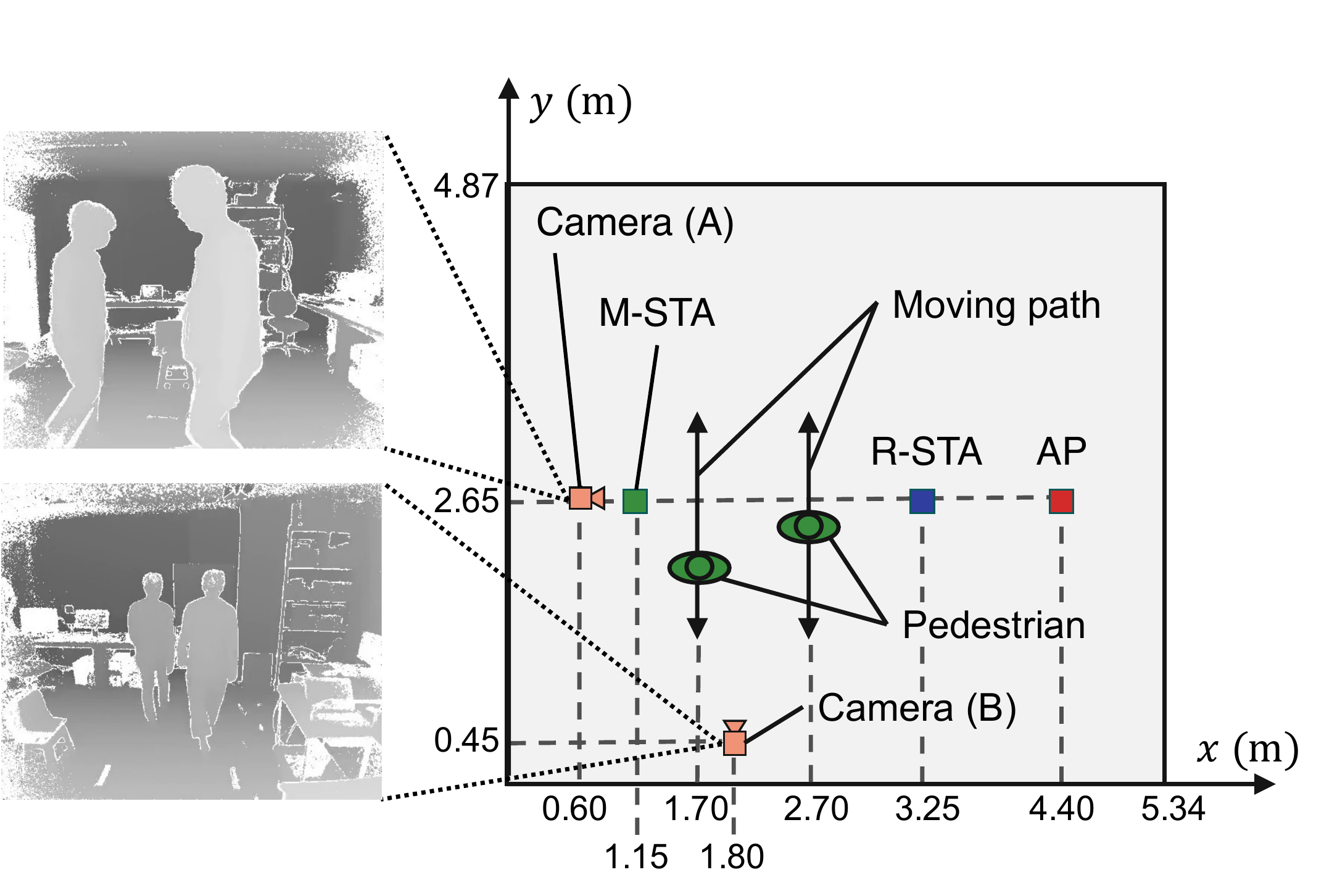}
    \caption{Experimental environment for obtaining original dataset (i.e., depth image dataset before converting to point cloud). The AP and the response STA (R-STA) communicate using mmWave, and RSSI is measured by the measurement STA (M-STA). The two left images are samples of depth images acquired by cameras A and B, respectively.}
    \label{fig:rssi_exp_env}
\end{figure}

The original dataset (i.e., depth image dataset before converting to point cloud) consists of a time series of pairs of mmWave RSSI and depth images acquired by a depth camera, with the time series measured at 30 frames per second (fps).
The experimental environment for obtaining this dataset is shown on \cref{fig:rssi_exp_env}.
The experimental environment included an AP, a response STA (R-STA) for communicating with the AP, a measurement STA (M-STA) for RSSI measurement, and a depth camera at position A or B.
A commercially available IEEE 802.11ad-compliant product was used as the AP.
The M-STA measures the RSSI of IEEE 802.11ad frames sent from the AP to the R-STA without being affected by the beamforming operation, which varies among IEEE 802.11ad products~\cite{koda:vtc2017hmm}.
We used Microsoft Kinect v2 as the depth camera, which uses infrared radiation and does not interfere with mmWave communications.
Details of the experimental equipment are summarized in \cref{tab:rssi_exp_equip}.

The experiment was conducted in a room where two pedestrians intermittently blocked the mmWave communication LOS path between the R-STA and the M-STA.
At the start of the experiment, the pedestrians stopped at the end of the moving path in \cref{fig:rssi_exp_env}.
The pedestrian traversed from one extremity of the movement path to the other at a steady pace, thereby obstructing the mmWave LOS path in the process. 
The walking speed was arbitrarily determined by the pedestrian at the onset of the movement, ensuring that they reached the opposing end of the path within a time span of 3 to 6\,s.
The procedure for determining the walking speed is executed each time the pedestrian arrives at the opposite end.
Upon reaching their respective opposite ends, the pedestrians initiated a pause, with a duration extending from 1 to 3\,s. 
This time frame also accounted for the period necessary to turn around. 
The pedestrians each repeated this cycle of walking and stopping. 
As a result, we created a dataset of non-periodic LOS blockage caused by pedestrians walking at various speeds.
The average interval between blockages per pedestrian was approximately 6\,s, resulting in an average of once every 4\,s LOS blockage because two people sometimes block the LOS simultaneously.
When LOS blockage occurs, the RSSI is attenuated by approximately 15\,dB, which is on the same level as the average value of the IEEE 802.11ad channel model, 13.4\,dB~\cite{alexander:channelmodels}.

The dataset acquired in the situation where the camera position is A or B, respectively, is referred to as dataset A or B, respectively.
Sample depth images of the two viewpoints in this dataset are shown on the left side of \cref{fig:rssi_exp_env}.
Measurements were taken for approximately 10\,min for camera positions A and B, and approximately 18,000 samples were measured.

Depth camera point clouds were generated by applying the process detailed in Appendix~B.
Specifically, from depth images with the shape of $(512, 424)$ where each pixel represents depth values from $0$ to $255$, normalized point clouds existing in the region $[0, 256)^3$ were generated.
In this paper, depth camera point clouds are the aforementioned normalized point clouds.
Generated depth camera point clouds contain information only in a specific direction because the depth camera observes a specific direction.

\begin{table}[!t]
  \caption{Experimental equipment on the depth camera point cloud and received signal strength dataset}
  \label{tab:rssi_exp_equip}
  \centering
  \begin{tabular}{ccc}
\toprule
                          & mmWave AP                  & Dell Wireless Dock D5000    \\
                          & mmWave R-STA                 & Dell Latitude E5540         \\
\multirow{2}{*}{Wireless} & M-STA antenna & Horn antenna, 24\,dBi       \\
                          & WLAN standard              & IEEE 802.11ad               \\
                          & Channel                    & 60.48\,GHz                  \\
                          & Bandwidth                  & 2.16\,GHz                   \\ \midrule
                          & Depth camera               & Microsoft Kinect v2 \\
Sensor                    & Depth image resolution           & 512$\times$424 \\
                          & Frame rate                 & 30\,fps                     \\ \bottomrule
\end{tabular}
\end{table}

\subsection{PREPROCESSING AND MACHINE LEARNING SETUPS}\label{ss:rssi_prepro_ml_setup}
The preprocessing described in Section~III-D is applied to the depth camera point cloud to generate the dataset corresponding to the time series voxel grids and RSSI values.
We experimentally determined the values of hyperparameters for preprocessing; the values are shown in \cref{tab:rssi_prepro_hypr}.
In particular, cuboid cropping removes the space in the large $z$-coordinate range and paddings the blank space so that the space size returns to $[0, 256)^3$ to remove the noise points in the foreground.
The number of points in the depth camera point cloud is fixed at 217,088, and subsequent preprocessing would be time-consuming if the number of points is not reduced.
Even if the link quality could be predicted 1000\,ms ahead, if the inference results are not available until earlier than 1000\,ms, the future prediction will be meaningless.
The reduction rate for random downsampling, $r_\mathrm{d}$, was experimentally determined to be 0.0921. 
This leaves approximately 20,000 points, thereby preventing the preprocessing latency from becoming excessively large.
The values of $n_\mathrm{o}$ and $r_\mathrm{o}$ were set to 20 and 2, respectively, with reference to Open3D~\cite{zhou:arxiv2018open3d} default values due to their balanced computation latency and noise reduction.
The value of $s_\mathrm{v}$ was set to 8 to be close to (40, 40), the input shape for the image-based method~\cite{nishio:jsac2019proactive}, to obtain voxel grids with a shape of (32, 32, 32).
In this configuration, the average latency for the preprocessing in our experiments using AMD EPYC 7542 CPU was 91\,ms, which is sufficiently short for predicting 500\,ms ahead and 1000\,ms ahead.
Examples of preprocessing of a depth image point cloud are shown on the left column of \cref{tab:example_prepro}.

These datasets were acquired at 30\,fps, and the model was input with a tensor that concatenated the voxel grids for 16 frames, corresponding to the last 500\,ms as well as our previous study~\cite{nishio:jsac2019proactive}.
The model predicts current and future values according to the system model shown in Section~III-A.
ML models predict RSSI 0\,ms and 500\,ms ahead, as well as our previous study~\cite{nishio:jsac2019proactive}.
In addition, ML models also predict 1000\,ms ahead.
Since delays on the order of 100\,ms occur when streaming UHD videos or VR content from a cloud server via the internet, communication control instructions can be given with time to spare by predicting 500\,ms or 1000\,ms ahead, also taking into account the time of link quality inference and communication control.
As described in Section~III-D, temporal difference labeling~\cite{nishio:jsac2019proactive} was used to create the time sequential dataset.

% \usepackage{multirow}
\begin{table}[t!]
\caption{Preprocessing hyperparameters for the depth camera point cloud and received signal strength dataset}
\label{tab:rssi_prepro_hypr}
\centering
\begin{tabular}{ccc} \toprule
Preprocessing phase            & Hyperparameter                       & Value                   \\ \midrule
\multirow{2.5}{*}{Cuboid cropping}                           & $(x_{\min},y_{\min},z_{\min})$          & (0, 0, 0)             \\ \cmidrule{2-3} 
                               & $(x_{\max}, y_{\max}, z_{\max})$        & (256, 256, 244)       \\ \midrule
Random downsampling            & $r_\mathrm{d}$                                & 0.0921                  \\ \midrule
\multirow{2.5}{*}{\begin{tabular}{c}Statistical\\outlier removal\end{tabular}}                    & $n_\mathrm{o}$                                & 20                      \\ \cmidrule{2-3} 
                               & $r_\mathrm{o}$                                & 2                       \\ \midrule
\multirow{2}{*}{Voxelization}  & $s_\mathrm{v}$                                & 8                       \\
                               & \multicolumn{2}{c}{(i.e., the voxel grid shape is (32, 32, 32))} \\ \midrule
\multirow{4.5}{*}{\begin{tabular}{c}Time series\\concatenation\end{tabular}} & $s$                                  & 16                      \\
                               & \multicolumn{2}{c}{(i.e., using latest 500\,ms time series data)}               \\ \cmidrule{2-3} 
                               & $k$                                  & 0, 15, 30               \\
                               & \multicolumn{2}{c}{(i.e., predicting 0, 500, 1000\,ms ahead)}               \\ \bottomrule
\end{tabular}
\end{table}

Three point cloud-based ML models, two NNs, Conv3D and ConvLSTM3D, and GBDT as proposed in Section~III-E, were used to predict RSSI.
We used Keras~\cite{keras} in TensorFlow~\cite{abadi:osdi2016tensorflow} as NN implementation.
The number of trainable parameters of Conv3D and ConvLSTM3D models for mmWave RSSI prediction were 253,121 and 148,353, respectively.
\cref{tab:hyperparameters} presents the hyperparameters used in the ML models. 
We adopted the MSE as the loss function for regression, as we did in our previous study~\cite{nishio:jsac2019proactive}.
We used default learning rate values in Keras~\cite{keras} and LightGBM~\cite{ke:nips2017lightgbm}.
For NN training, we used mean-squared-error (MSE) for the loss function and utilized ReduceLROnPlateau, which is a learning rate scheduler that reduces the learning rate by a factor of four if the validation loss did not improve for two epochs.
In our experiments using the NVIDIA Quadro RTX 6000 GPU, the average computational latencies for Conv3D, ConvLSTM3D, and GBDT were 49\,ms, 81\,ms, and 23\,ms, respectively.
In addition to these three point cloud-based models, two non-point cloud-based methods were prepared for comparative evaluation: an RSSI time series-based method and a depth image-based method.
The RSSI time series-based method uses only the time series of the previous RSSI as features and predicts RSSI values using the GBDT algorithm.
The depth image-based method is the same as \cite{nishio:jsac2019proactive}, and depth images before conversion to point clouds are used as the input feature.
These two methods are the same as point cloud-based methods in that they use the latest 500\,ms data to predict RSSI.

\begin{table}[t!]
    \caption{Hyperparameters for ML model training}
    \centering
    \scalebox{0.95}{
    \begin{tabular}{ccc}
    \toprule
    Algorithm & Hyperparameter & Value \\ \midrule
                          & Loss function                  & MSE  \\
            Conv3D          & Optimizer                      &  Adam~\cite{kingma:iclr2015adam}  \\
            \&          & Learning rate                  & 0.001 \\
            ConvLSTM3D          & Learning rate scheduler & ReduceLROnPlateau\\
                          & Early stopping epochs          & 5    \\ \midrule
    \multirow{4}{*}{GBDT} & Objective function            & RMSE  \\
                          & Number of leaves               & 31    \\
                          & Learning rate                  & 0.1   \\
                          & Early stopping boosting rounds & 20    \\ \bottomrule
    \end{tabular}
    }
    \label{tab:hyperparameters}
\end{table}

We evaluated our method using datasets A and B.
These two datasets both consist of time series data of approximately 10\,min.
We used the first 60\% for training the ML model, the next 20\% for validation during model training, and the last 20\% for holdout validation used for evaluation.
Cross-validation was not used to prevent data leakage of time series data and unbalanced or small training data volume.

\subsection{EXPERIMENTAL RESULTS}\label{ss:rssi_exp_results}
\begin{figure*}[t!]
    \centering
    \includegraphics[width=180mm]{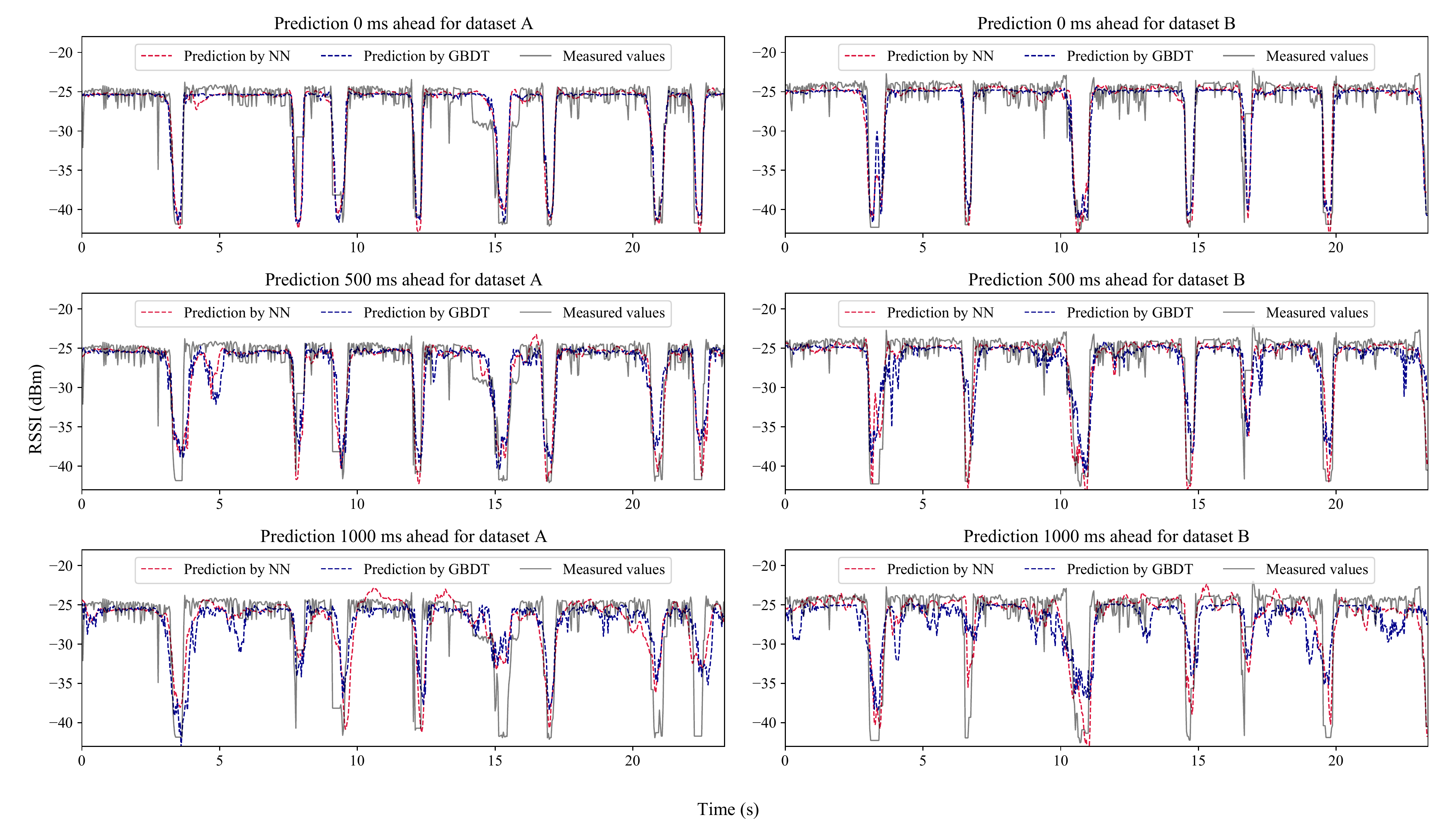}
    \vspace{-6mm}
    \caption{Predicted and measured RSSI values}
    \label{fig:rssi_pred}
\end{figure*}

We first conducted a qualitative evaluation of the link quality prediction method by plotting the measured and predicted RSSI values, as shown in \cref{fig:rssi_pred}.
The RSSI values were predicted from time series voxel grids using Conv3D and GBDT, presented in Section~III-E.
The left and right columns are the results for datasets A and B, that is, camera positions A and B, respectively.
The first row is 0\,ms ahead, i.e., the current prediction, while the second and third rows are future predictions 500\,ms and 1000\,ms ahead, respectively.
The measured RSSI values are significantly attenuated when pedestrians block the LOS of mmWave communications.
Correspondingly, the predicted RSSI values also attenuate significantly, suggesting that the model can predict blockage.
For camera positions A and B, the measured and predicted values appear to match in both cases.
Comparing the two models, Conv3D and GBDT, Conv3D provided a better match better, albeit slightly.
Based on comparisons of predictions 0, 500, and 1000\,ms ahead, the discrepancy between the prediction and the actual measurement is greater when the prediction is further ahead in time.
In particular, this tendency can be observed for large attenuations, such as LOS blockage, and may occur owing to the difficulty in predicting the time further ahead.

Next, we quantitatively evaluated link quality prediction errors.
The RMSE was used to evaluate the prediction accuracy, considering the accuracy of the prediction during LOS blockage, because the RMSE reflects the effects of outliers.
The RMSE values in RSSI prediction using five methods are listed in \cref{tab:rssi_result}.
The RSSI-based and depth image-based methods are for comparison with point cloud-based methods as described in Section~IV-B.
In the proposed point cloud-based method, three models, namely Conv3D, ConvLSTM3D, and GBDT, were compared.
Each model inputs 16 frames, corresponding to the last 500\,ms, as input feature and predicts the RSSI values.
For Conv3D and ConvLSTM3D, \cref{tab:rssi_result} shows the average RMSE from four trials using different initial weights, taking into account the initial weight dependence of NN models. 
Conversely, since the difference in performance due to different random seeds was negligible in the methods using GBDT, only the result from one trial is shown.

% \usepackage{multirow}
\begin{table}[t]
\centering
\caption{RMSE values of RSSI prediction}
\scalebox{0.92}{
\begin{tabular}{ccc|cc} \toprule
\multirow{2}{*}{Ahead}    & \multirow{2}{*}{Method}     & \multirow{2}{*}{Model} & \multicolumn{2}{c}{RMSE (dB)} \\ 
                          &                              &                         & Dataset A     & Dataset B     \\ \midrule
\multirow{6}{*}{0\,ms}    & RSSI-based                    & GBDT                    & ---           & ---           \\ \cmidrule{2-5}
                          & Depth image-based                  & NN                      & \textbf{2.319}         & 2.612         \\ \cmidrule{2-5} 
                          & \multirow{3}{*}{Point cloud-based} & Conv3D                      & 2.338         & 2.374         \\
                          &                              & ConvLSTM3D                    & 2.492         & 2.480         \\
                          &                              & GBDT                    & 2.338         & \textbf{2.324}         \\ \midrule 
\multirow{6}{*}{500\,ms}  & RSSI-based                    & GBDT                    & 5.024         & 4.973         \\ \cmidrule{2-5} 
                          & Depth image-based                  & NN                      & 3.284         & 3.594         \\ \cmidrule{2-5} 
                          & \multirow{3}{*}{Point cloud-based} & Conv3D                      & \textbf{3.236}         & \textbf{3.333}         \\
                          &                              & ConvLSTM3D                    & 3.388         & 3.501         \\
                          &                              & GBDT                    & 3.428         & 3.496         \\ \midrule 
\multirow{6}{*}{1000\,ms} & RSSI-based                    & GBDT                    & 5.068         & 5.053         \\ \cmidrule{2-5} 
                          & Depth image-based                  & NN                      & 4.363         & 4.262         \\ \cmidrule{2-5} 
                          & \multirow{3}{*}{Point cloud-based} & Conv3D                      & \textbf{3.992}         & \textbf{3.754}         \\
                          &                              & ConvLSTM3D                    & 4.342         & 4.286         \\
                          &                              & GBDT                    & 4.311         & 4.294         \\ \bottomrule
\end{tabular}
}
\label{tab:rssi_result}
\end{table}

In \cref{tab:rssi_result}, the RSSI time series-based method exhibits significantly larger error values compared to the other methods, and this is likely due to the unpredictability of LOS path blockage. 
These results suggest that accurately predicting LOS blockage is a challenging task that cannot be achieved solely by considering the previous link quality values.
The point cloud-based method could predict 0\,ms and 500\,ms ahead with approximately 2.3\,dB and 3.2\,dB, respectively, with almost the same errors, compared with those of existing depth image-based methods.
Furthermore, for the prediction of 1000\,ms ahead, Conv3D had an error of approximately 10\% smaller than the depth image-based method and ConvLSTM3D and the GBDT model of the point cloud-based method.
This might be because the convolution layer in the Conv3D enables accurate spatio-temporal understanding.
Accordingly, we conclude that the point cloud-based method can predict LOS path blockage as well as or better than the depth image-based method.

\begin{figure}[t!]
    \centering
    \includegraphics[width=85mm]{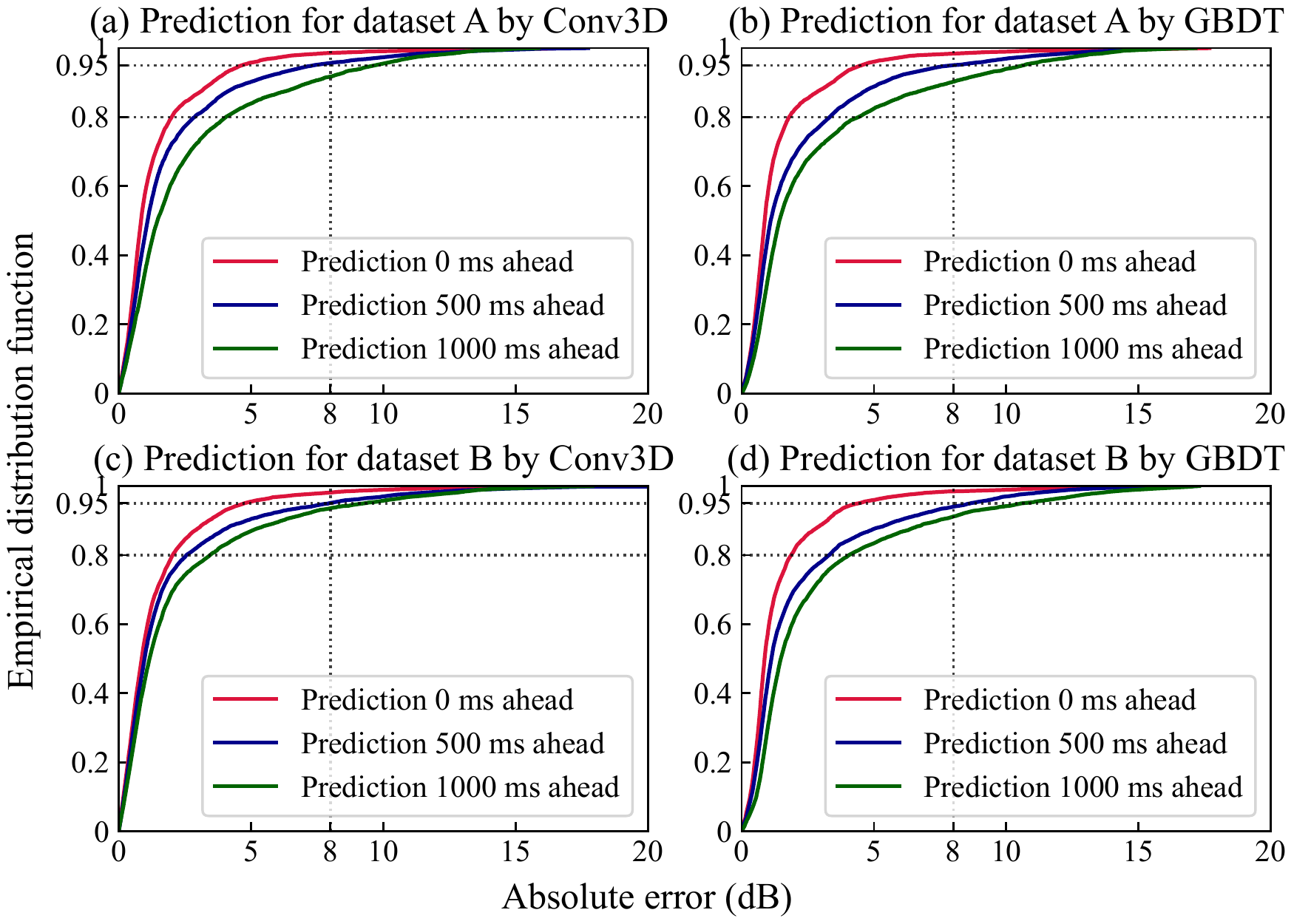}
    \caption{The empirical distribution function of absolute errors for RSSI prediction. Two horizontal dotted lines represent 0.8 and 0.95.}
    \label{fig:rssi_edfa}
\end{figure}

The empirical distribution function of absolute errors is depicted in \cref{fig:rssi_edfa}, where it can be observed that the 80th percentile absolute error value is less than 5\,dB for all the predictions at 0, 500, and 1000\,ms ahead, indicating that the errors are concentrated within 5 dB or less.
Additionally, for all predictions at 0, 500, and 1000\,ms ahead, the 95th percentile absolute error value was less than 12\,dB, indicating that there were only a few errors above 12\,dB.

During LOS communication in this experiment, the received power remains around $-$25\,dBm with little variation. 
In contrast, the received power significantly attenuated during LOS blockage caused by pedestrians. 
When the attenuation exceeds 8\,dB from the median received power during LOS communication, $-$25.05\,dBm, it is considered as a blockage. 
The percentage of blockage time is 13.0\% and 11.9\% for datasets A and B, respectively. 
However, as shown in \cref{fig:rssi_edfa}, the percentage of errors greater than 8\,dB is very small, below 5\% for the 500\,ms prediction and below 10\% for the 1000\,ms prediction. 
This suggests that most of the blockages could be predicted and the model is unlikely to fail to predict complete blockage. 
Furthermore, as shown in \cref{fig:rssi_pred}, the errors include cases where the starting time of attenuation due to blockage was correctly predicted but the amount of attenuation was incorrect, as well as cases where the amount of attenuation was correct, but there was a time shift in the start/end of attenuation due to blockage.
Therefore, we conclude that our point cloud-based link quality prediction method can predict RSSI attenuation due to LOS blockage.

\section{EXPERIMENTAL EVALUATION USING LIDAR POINT CLOUD AND MILLIMETER-WAVE THROUGHPUT DATASET}\label{s:thrp}
\subsection{EXPERIMENT FOR OBTAINING DATASET}\label{ss:thrp_exp_dataset}
We newly conducted a mmWave communication experiment to obtain a LiDAR point cloud dataset labeled with throughput values of mmWave communication to evaluate our proposed method.
In this experiment, we used mechanical rotation LiDAR, which scans spatial information in all horizontal directions.
Hence, 360\textdegree\, point clouds can be obtained from mechanical rotation LiDAR, unlike depth camera point clouds which only contain spatial information for a specific horizontal angle.
Specifically, we used Velodyne VLP-16 LiDAR, a widely used mechanical rotation LiDAR product, that can acquire information on vertical angles from $-$15\textdegree\, to $+$15\textdegree.
Velodyne VLP-16 LiDAR uses near-infrared light with a wavelength of 905\,nm~\cite{vlp}; thus, LiDAR does not interfere with mmWave communications.

The transmission control protocol (TCP) throughput of IEEE 802.11ad mmWave communications was used for the link quality indicator.
TCP throughput values are influenced by many factors, such as beamforming, the transmission rate of the AP, and congestion control of the TCP.
During the experiment, beamforming, rate control, and congestion control were frequently activated due to LOS blockage, resulting in increased dynamics in the throughput values.

\begin{figure}[t!]
    \centering
    \includegraphics[width=85mm]{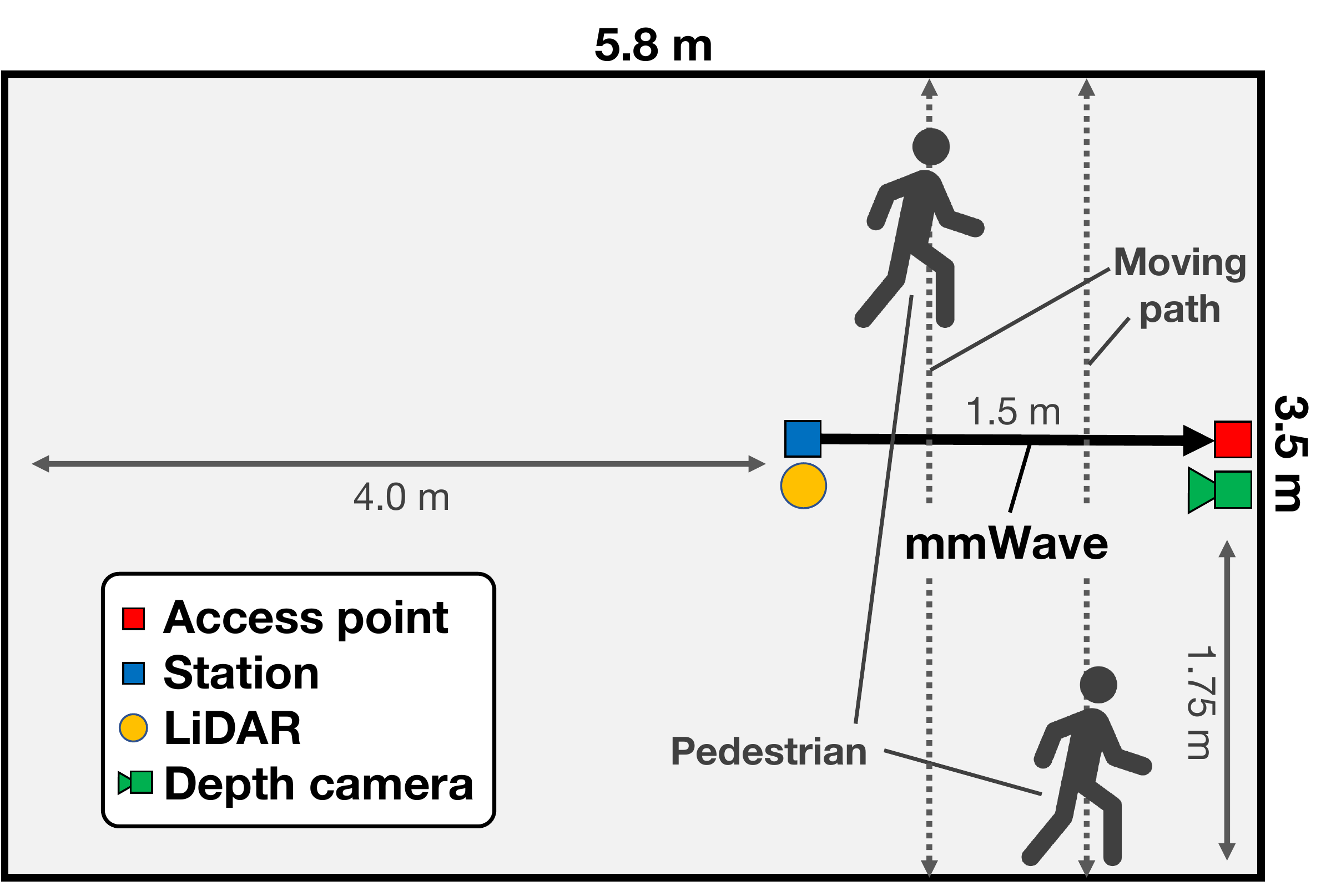}
    \caption{Experimental environment for obtaining LiDAR point cloud and mmWave TCP throughput dataset}
    \label{fig:thrp_exp_env}
\end{figure}

\begin{figure}
    \centering
    \includegraphics[width=85mm]{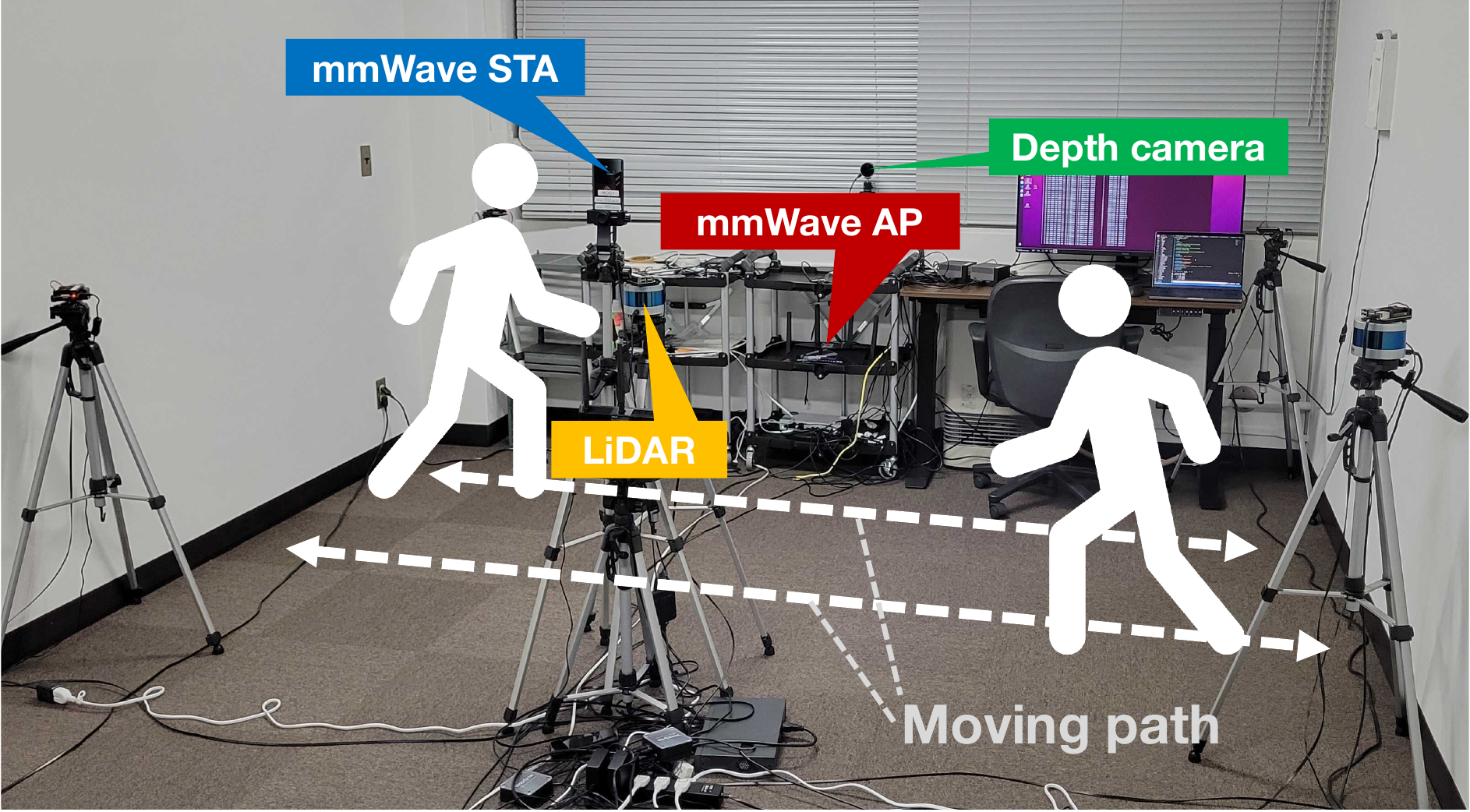}
    \caption{Experimental setup to obtain mmWave throughput and LiDAR point cloud. The commercially available IEEE 802.11ad-based 60\,GHz communication devices and Velodyne VLP-16 LiDAR were used.}
    \label{fig:thrp_exp_env_photo}
\end{figure}

Figs.~\ref{fig:thrp_exp_env} and \ref{fig:thrp_exp_env_photo} depict the indoor experimental setup and a photograph of the environment, respectively. 
\cref{tab:thrp_exp_equip} provides a comprehensive list of the equipment installed in the environment and their respective settings.
Both AP and STA used commercially available products. 
Furthermore, a smartphone was used for the STA to make the experimental environment more practical.
mmWave communications between the AP and STA used 60\,GHz IEEE 802.11ad WLAN.
We used iperf~\cite{tirumala1999iperf} to generate uplink TCP traffic, i.e., from the STA to the AP.
Throughput was calculated using Wireshark~\cite{wireshark} as the sum of the length of packets obtained by packet capture by tcpdump~\cite{tcpdump} in the last 100\,ms.

Two pedestrians intermittently blocked the mmWave LOS path between the AP and the STA. Their moving paths are visually represented in  Figs.~\ref{fig:thrp_exp_env} and \ref{fig:thrp_exp_env_photo}. 
The pedestrian movement patterns were consistent with those detailed in Section IV-A. 
Their walking speed was adequate for traversing the designated path within a range of 3 to 6\,s, while the duration for pausing at the end of the path ranged from 1 to 3\,s.
As a result, two pedestrians blocked the mmWave LOS path approximately once in 4\,s on average.

The experimental environment was observed using a LiDAR device installed at the center of the room.
LiDAR continuously provided point clouds of the experimental environment in a 3D Cartesian coordinate system with the origin located at the position of the LiDAR device.
An example of the obtained LiDAR point cloud data is shown in the figure on the right column of \cref{tab:example_prepro}.
The average number of points in one LiDAR point cloud was approximately 28,000.
In addition to LiDAR, a camera was set up next to the AP to compare with point clouds.
The point clouds, images, and throughput values were measured at a rate of 10\,fps for 30\,min, resulting in a dataset with 18,000 samples.

\begin{table}[!t]
  \caption{Experimental equipment and settings on the LiDAR point cloud and throughput dataset.}
  \label{tab:thrp_exp_equip}
  \centering
      \begin{tabular}{ccc}
\toprule
\multirow{5}{*}{Wireless} & mmWave AP                      & NETGEAR Nighthawk X10           \\
                          & mmWave STA                     & ASUS ROG Phone                  \\
                          & WLAN standard                  & IEEE 802.11ad                   \\
                          & Channel frequency              & 60.48\,GHz                      \\
                          & Channel bandwidth              & 2.16\,GHz                       \\ \midrule
\multirow{2}{*}{iperf~\cite{tirumala1999iperf}}    & Transport layer protocol & TCP                             \\
                          & Traffic direction        & Uplink (i.e., from STA to AP) \\ \midrule
\multirow{3}{*}{Sensor}   & LiDAR                          & Velodyne VLP-16                 \\
                          & Depth camera                         & Intel RealSense L515            \\
                          & Frame rate                     & 10\,fps                         \\ \bottomrule
\end{tabular}
\end{table}

\subsection{PREPROCESSING AND MACHINE LEARNING SETUP}
Our proposed point cloud-based link quality prediction method was evaluated using LiDAR point clouds and throughput values obtained in the aforementioned experiment.
As with Section~IV-B, the preprocessing described in Section~III-D was applied to the LiDAR point cloud to generate a voxel grid time series data and throughput corresponding dataset.
We determined values of preprocessing hyperparameters, shown in \cref{tab:thrp_prepro_hypr}.
First, points outside the cuboid region with vertices (-5,-5,-5) and (5,5,5), which are obviously noise points outside the room, were removed because the size of the indoor environment used in this experiment was 5.8\,m~\texttimes~3.5\,m.
In the experimental environment, the average number of LiDAR point cloud points was 28,826, which allowed subsequent processing to be computed without losing the meaning of future predictions.
Therefore, random downsampling was not performed and $r_\mathrm{d}$ was set to 1.
As with Section~IV-B, the values of $n_\mathrm{o}$ and $r_\mathrm{o}$ were set to 20 and 2, respectively, with reference to Open3D~\cite{zhou:arxiv2018open3d} default values.
In the voxelization, $s_\mathrm{v}$ was set to 0.2\,m in order to divide as roughly as possible while preserving the human shape.
The difference from Section IV-B is that cuboid cropping is performed without padding the blank space after cropping, resulting in a narrower space.
This is to prevent the large useless calculations of the following process caused by padding the blank space.

We used three point cloud-based ML models, two NNs, Conv3D and ConvLSTM3D, and GBDT as proposed in Section~III-E to predict throughput.
The number of trainable parameters of Conv3D and ConvLSTM3D models for mmWave throughput prediction were 232,257 and 141,441, respectively.
The dataset consisted of 18,000 samples of 30\,min at 10\,fps. 
The first 60\% was used to train the ML model, the next 20\% was used for validation during model training, and the last 20\% was used for holdout validation to evaluate our link quality prediction method.
The dataset was obtained at 10\,fps, and features were input to the model for the six previous frames, corresponding to the last 500\,ms as well as our previous study~\cite{nishio:jsac2019proactive}.
As with Section~IV-B, ML models predict 0\,ms ahead, 500\,ms ahead, and 1000\,ms ahead. 
We used temporal difference labeling~\cite{nishio:jsac2019proactive} to create the dataset for training the ML model.
We compared the performance of point cloud-based methods with the image-based method and the throughput-based method that replaces RSSI with throughput in the RSSI-based method used in Section~IV.

% \usepackage{multirow}
\begin{table}[t!]
\caption{Preprocessing hyperparameters for LiDAR point cloud and throughput dataset}
\label{tab:thrp_prepro_hypr}
\centering
\begin{tabular}{ccc} \toprule
Preprocessing phase            & Hyperparameter                       & Value                   \\ \midrule
\multirow{2.5}{*}{Cuboid cropping}                    & $(x_{\min},y_{\min},z_{\min})$          & (-5\,m, -5\,m, -5\,m)             \\ \cmidrule{2-3} 
                               & $(x_{\max}, y_{\max}, z_{\max})$        & (5\,m, 5\,m, 5\,m)       \\ \midrule
Random downsampling            & $r_\mathrm{d}$                                & 1                  \\ \midrule
\multirow{2.5}{*}{\begin{tabular}{c}Statistical\\outlier removal\end{tabular}}                    & $n_\mathrm{o}$                                & 20                      \\ \cmidrule{2-3} 
                               & $r_\mathrm{o}$                                & 2                       \\ \midrule
\multirow{2}{*}{Voxelization}  & $s_\mathrm{v}$                                & 0.2\,m                       \\
                               & \multicolumn{2}{c}{(i.e., the voxel grid shape is (23, 33, 10))} \\ \midrule
\multirow{4.5}{*}{\begin{tabular}{c}Time series\\concatenation\end{tabular}} & $s$                                  & 6                      \\
                               & \multicolumn{2}{c}{(i.e., using latest 500\,ms time series data)}               \\ \cmidrule{2-3} 
                               & $k$                                  & 0, 5, 10               \\
                               & \multicolumn{2}{c}{(i.e., predicting 0, 500, 1000\,ms ahead)}               \\ \bottomrule
\end{tabular}
\end{table}

\subsection{EXPERIMENTAL RESULTS}
\begin{figure}[t!]
    \centering
    \includegraphics[width=85mm]{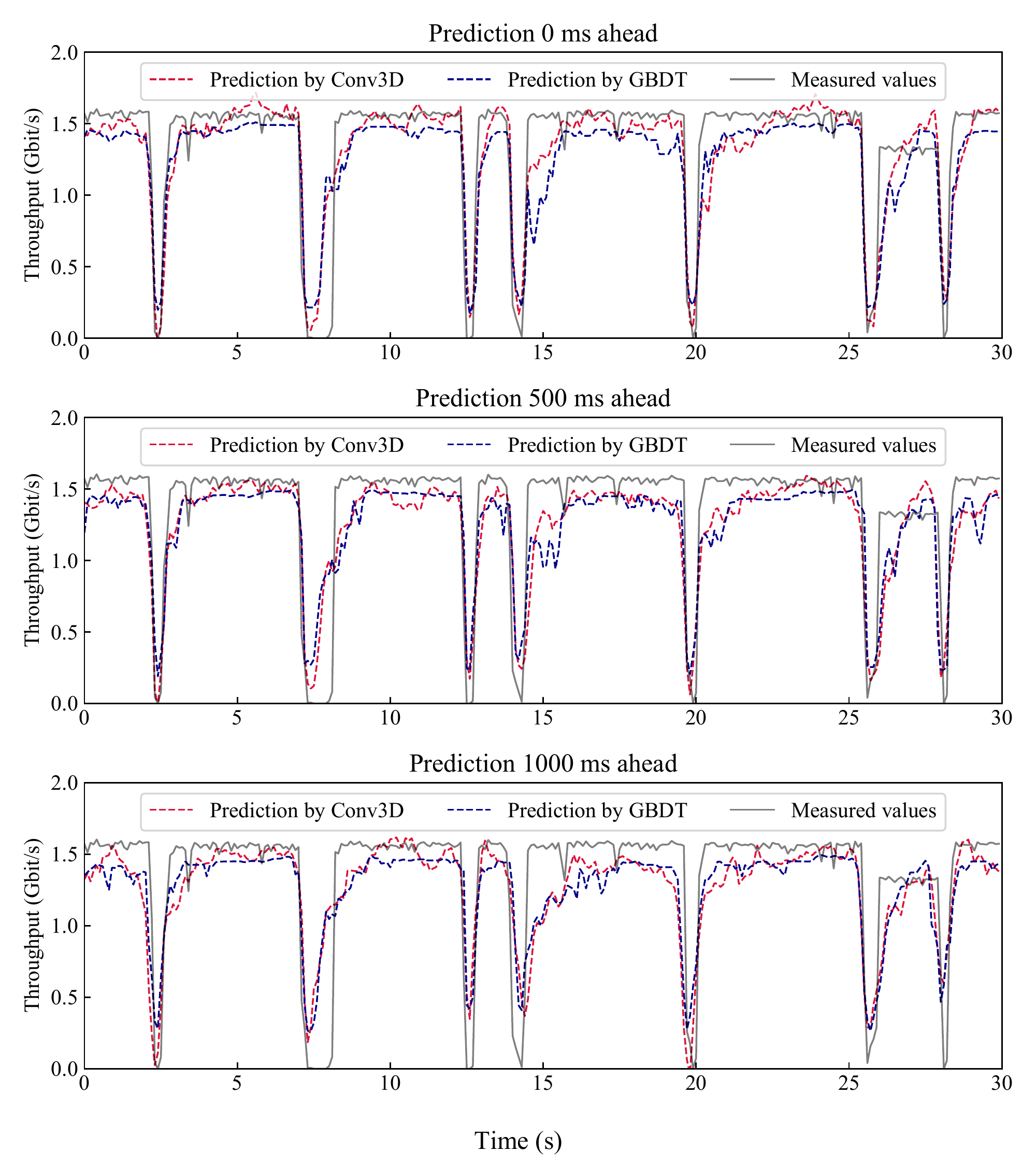}
    \caption{Predicted and measured throughput values}
    \label{fig:thrp_pred}
\end{figure}

\cref{fig:thrp_pred} displays examples of measured and predicted throughput using the point cloud-based methods with Conv3D and GBDT models.
The throughput value of mmWave during LOS communication was approximately 1.6\,Gbit/s.
When the LOS path was blocked by a pedestrian, the throughput value attenuated to 0\,Gbit/s.
When the measured values were significantly attenuated, the predicted values were significantly attenuated simultaneously.
Two consecutive throughput values degradations occurred around 13\,s were also able to be predicted.
The point cloud-based methods demonstrated the capability to accurately predict the occurrence of mmWave LOS blockage caused by the two pedestrians. 
However, the proposed method was not always able to accurately predict the throughput at the end of the blockage, as shown around 7\,s. 
This is because, unlike RSSI, the throughput value is determined by a complex interplay of factors, such as beamforming, the transmission rate of the AP, and congestion control of TCP, making it challenging to predict when the throughput value fully recovers only from the spatial information of point clouds.

Comparing the predictions at 0, 500, and 1000\,ms ahead, the predicted values do not often deteriorate to 0\,Gbit/s in the predictions at more future timesteps ahead.
This could be attributed to the difficulty in predicting future blockages, and the uncertainty of the model in predicting their occurrence. Thus, the model outputs a prediction with the expected value of RMSE to be as small as possible in anticipation of the case where no blockage occurs.
Similarly, the slightly lower predicted throughput during LOS communication could be due to the degradation in throughput after the blockage, caused by beamforming, transmission rate control, and congestion control, as observed in the interval between 25\,s and 28\,s in \cref{fig:thrp_pred}. 
Looking at the predicted results during blockage, the Conv3D model appears to be more in agreement with the actual values compared to the GBDT model. This is thought to be due to the use of 3D convolution in the Conv3D model, which is more suitable for capturing spatio-temporal features.

\cref{tab:thrp_result} presents the RMSE values for the throughput prediction, where we compare the performance of the proposed point cloud-based methods using three different ML models with the throughput-based and depth image-based methods.
Predictions of the throughput-based method have larger error values than those of other methods, which indicate that blockage could not be predicted.
Predicting throughput values accurately in a dynamically changing mmWave communication environment using only the time series of throughput is a challenging task, similar to RSSI prediction. 
The point cloud-based methods employing Conv3D, ConvLSTM3D, and GBDT outperform the throughput-based method and achieve performance that is comparable or even superior to the depth image-based method, particularly when predicting blockages further ahead in time.
This is likely due to the fact that the point cloud data acquired from LiDAR can provide information from a wider field of view, which allows for more accurate predictions of future conditions.

\begin{table}[t]
\centering
\caption{RMSE values of throughput prediction}
\label{tab:thrp_result}
\begin{tabular}{ccc|c} \toprule
Ahead                     & Method                      & Model & RMSE (Gbit/s) \\ \midrule
\multirow{6}{*}{0\,ms}    & Throughput-based              & GBDT   & ---           \\ \cmidrule{2-4}
                          & Depth image-based                  & NN     & 0.2771        \\ \cmidrule{2-4}
                          & \multirow{3}{*}{Point cloud-based} & Conv3D     & \textbf{0.2747}        \\
                          &                              & ConvLSTM3D   & 0.2787        \\
                          &                              & GBDT   & 0.2767        \\ \midrule
\multirow{6}{*}{500\,ms}  & Throughput-based              & GBDT   & 0.4435        \\ \cmidrule{2-4}
                          & Depth image-based                  & NN     & 0.3178        \\ \cmidrule{2-4}
                          & \multirow{3}{*}{Point cloud-based} & Conv3D     & \textbf{0.2909}        \\
                          &                              & ConvLSTM3D   & 0.3139        \\
                          &                              & GBDT   & 0.2924        \\ \midrule
\multirow{6}{*}{1000\,ms} & Throughput-based              & GBDT   & 0.4497        \\ \cmidrule{2-4}
                          & Depth image-based                  & NN     & 0.3756        \\ \cmidrule{2-4}
                          & \multirow{3}{*}{Point cloud-based} & Conv3D     & 0.3200        \\
                          &                              & ConvLSTM3D   & 0.3399        \\
                          &                              & GBDT   & \textbf{0.3133}        \\ \bottomrule
\end{tabular}
\end{table}

\begin{figure}[t!]
    \centering
    \includegraphics[width=85mm]{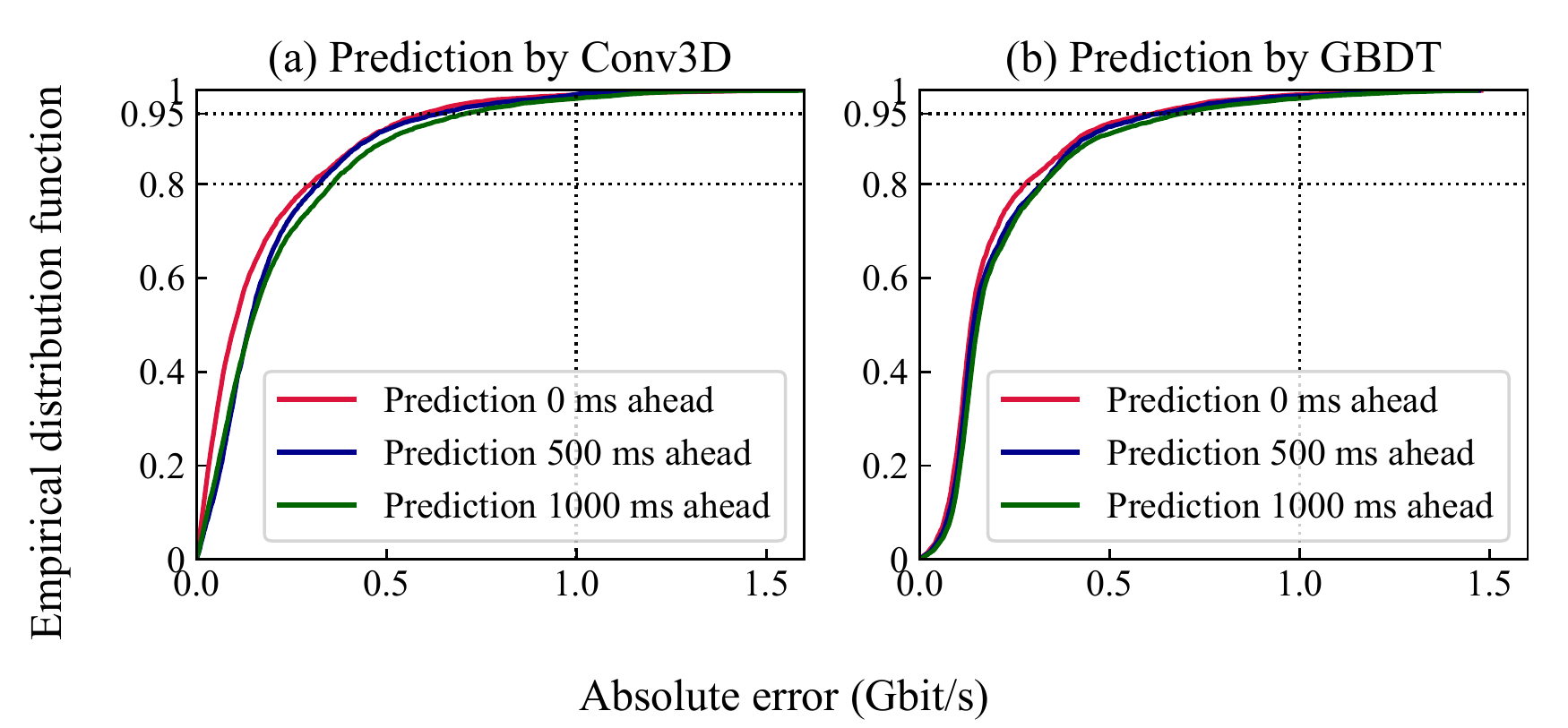}
    \caption{The empirical distribution function of absolute errors for throughput prediction. Two horizontal dotted lines represent 0.8 and 0.95.}
    \label{fig:thrp_edfa}
\end{figure}

The empirical distribution function of absolute prediction errors is presented in \cref{fig:thrp_edfa}. 
\cref{fig:thrp_edfa} indicates that the 80th percentile absolute errors ranged from 0.3 to 0.4\,Gbit/s for predictions at 0, 500, and 1000\,ms ahead. 
These errors were primarily due to fluctuations in throughput during LOS communication, which were approximately 0.3\,Gbit/s. 
In contrast, the percentage of errors exceeding 1\,Gbit/s was less than 1\%. 
During LOS communication, the throughput value was around 1.6\,Gbit/s, while during LOS blockage, the throughput often dropped to 0\,Gbit/s. 
Consequently, the attenuation due to blockage was 1.6\,Gbit/s, and if a blockage was not fully predicted, an absolute error of 1.6\,Gbit/s occurred. 
However, although the periods when throughput dropped by 1\,Gbit/s or more accounted for 11.2\% of the total, the proportion of prediction errors exceeding 1\,Gbit/s remained below 1\%, which is relatively small in comparison. 
Thus, we conclude that our point cloud-based method can effectively predict throughput values for both blockage and LOS communication scenarios.

\section{DISCUSSION}
In this section, we will discuss the remaining challenges and future research directions for point-cloud-based link quality prediction. 
Specifically, in Section~VI-A, we will delve into the ML algorithms in detail, while other challenges will be discussed in Section~VI-B.

\subsection{MACHINE LEARNING ALGORITHMS}
This study aims to demonstrate the feasibility of mmWave link quality prediction from point clouds. 
As such, we have adopted a simple approach that involves converting point clouds into voxel grid data and applying well-established algorithms such as NN with 3D convolution layers and 3D convolutional LSTM layers, and GBDT. 
One of the advantages of this method is that it uses voxel grids as the data format, which can be easily extended and applied to image-based ML algorithms while retaining the 3D structure. 
However, other approaches can also be considered. 
These approaches can be categorized into three distinct categories based on the format of input features: point clouds, 3D data representations other than point clouds, and hand-made features.

As mentioned in Section~III-C, the most straightforward approach is to use NN models designed for the point cloud, such as PointNet~\cite{qi:cvpr2017pointnet, qi2017pointnet++} and VoteNet~\cite{qi2019vote}, which can directly input point clouds and extract features from the points. 
Our preliminary experiments leveraged the existing models (i.e., PointNet and VoteNet) for learning the direct mapping from point clouds to RSSI. 
However, they failed to predict the large attenuation of link quality induced by LOS blockage. 
This is because the models' target task and required characteristics differ from ours. 
Generally, these existing models are, used for 3D object detection or segmentation, in which the translation invariant convolution in PointNet properly functions to detect objects regardless of their positions. 
However, the positions of objects are crucial in link quality prediction since mmWave communications are significantly affected by the mobility of obstacles and the positions of reflectors.
Designing a NN architecture that is suitable for directly predicting link quality from point clouds can be a new challenge in the field of vision-wireless ML and has the potential to improve prediction accuracy. 
One possible approach is to incorporate attention mechanisms, which have achieved success in natural language processing and computer vision fields, into point clouds or voxel grids~\cite{vaswani:nips2017attention, zhao2021point, mao2021voxel}. 
Attention mechanisms can extract information on the spatial locations of objects that affect link quality and dynamically determine the locations and objects of interest. 
Thus, attention mechanisms can potentially improve prediction accuracy by focusing on the relevant areas and objects.

Various data formats, such as meshes, octrees~\cite{rusu:icra2011pcl}, and implicit function representations using NN models~\cite{eslami:science2018neural, mildenhall2020nerf}, can be used to represent 3D spaces besides point clouds. 
However, the conversion between these formats may result in the loss of shape information. 
For instance, projecting the point cloud onto a 2D image~\cite{simony2018complex, ali2018yolo3d}, transforming the point cloud into a pseudo-image~\cite{lang2019pointpillars, yin2021center}, or converting the point cloud into sparse voxel grids~\cite{choy20194minkowski} are possible. 
Although most of these transformations have been proposed for robot control and autonomous driving, they are also relevant to the link quality prediction task, which involves recognizing and tracking objects in 3D space. 
By taking advantage of these data format conversions and exploring new ML models specifically designed for point cloud-based link quality prediction, we may be able to achieve higher prediction accuracy while reducing the necessary training data volume.

One alternative approach to improving link quality prediction is to enhance the feature engineering method in the preprocessing unit. Instead of relying solely on end-to-end ML inference with point clouds as inputs and link quality predictions as outputs, we can extract effective information from the point clouds using rule-based feature engineering techniques. This information can then be used as inputs to ML models for link quality prediction.
For instance, some previous studies~\cite{wu:wcnc2022lidar, marasinghe:gcw2021lidar, zhang:infocom2022vision} have extracted or converted point clouds to bounding boxes and heatmaps. 
Hand-crafted feature engineering can enhance the interpretability of the predictions and decrease the input dimensions, which, in turn, reduces the model complexity. 
However, as demonstrated in recent computer vision and NLP tasks, these feature engineering methods often fall short in prediction accuracy and generalization compared to state-of-the-art end-to-end ML. 
Therefore, developing feature extraction methods that can efficiently reduce the dimensionality of data while achieving high explainability, prediction accuracy, and generalization is a new research challenge.

Adapting to data drift caused by environmental changes or channel mobility also remains a challenge for link quality prediction. 
In ML, accuracy decreases when there is a domain discrepancy between the data used for training and the test data. 
Although the generalization capability of ML can absorb short-term and minor fluctuations caused by the randomness of the wireless channel, changes in the wireless channel and point clouds caused by variations in the positions of communication devices and point cloud sensors, as well as the configuration of surrounding furniture, can lead to data drift and decreased prediction accuracy. 
Our previous work~\cite{ohta:vtc2022mill} has demonstrated accuracy degradation in the prediction system when the test set was obtained several days after the training set, even though the experimental setup was not significantly changed.
To mitigate this issue, our prior research has shown that fine-tuning the model with a small amount of newly acquired data in the current environment can be effective~\cite{ohta:vtc2022mill}.
However, there is a need to investigate more efficient fine-tuning techniques, particularly in terms of optimizing the types and amounts of data used, the frequency of fine-tuning implementation, and exploring methods to prevent forgetting previously learned information due to excessive fine-tuning.

\subsection{ADVANCED RESEARCH TOPICS}\label{ss:discuss_future}
We here discuss more advanced research challenges, namely predicting the more detailed information on mmWave communication channels and the potential applications of mmWave communication signals to visual sensing tasks.

We believe that point clouds have the potential to enable higher-dimensional measures of wireless link quality prediction, specifically channel state information (CSI), beyond RSSI or throughput. 
In mmWave communications with 5G or IEEE 802.11ay, multi-input multi-output (MIMO) technology is used to achieve high speed and capacity through spatial multiplexing, as well as high-quality communication through diversity effects. 
As a result, accurate CSI estimation has become increasingly crucial in mmWave communications. 
Recently, a study demonstrated that CSI can be predicted from depth images in the 2.4\,GHz band~\cite{veni}, suggesting that point clouds, which provide a 3D representation of space, may also enable CSI prediction in the 60\,GHz band.

Another interesting direction for future research is to explore the inverse transformation of this study, namely, estimating spatial information from mmWave communication signals. 
It has already been demonstrated in the 5\,GHz band that images captured by a camera installed in the same room can be estimated based on Wi-Fi channel information, such as CSI~\cite{csi2image} or RSSI~\cite{rssi2image}, in indoor environments. 
Furthermore, it has been shown that image generation of LOS communication paths can be achieved using 60\,GHz band RSSI time series~\cite{nishio:csm2021when}. 
Similar to prior studies that generate images representing indoor conditions, it is anticipated that point clouds of indoor spaces can be derived from mmWave communication channel information. 
In particular, there is ample opportunity to investigate effective features and model architectures for indoor point cloud generation.
\section{CONCLUSION}
In this paper, we demonstrated the potential of point clouds as an alternative to camera images for proactive link quality prediction in mmWave communications. 
We devised a prediction framework for the point cloud-based link quality prediction, which incorporates preprocessing methods for point clouds, including cropping, downsampling, outlier removal, voxelization, time series concatenation, and labeling. 
We then examined three ML models—Conv3D, ConvLSTM3D, and GBDT—that can extract spatio-temporal features from time series voxel grids. 
Our point cloud-based link quality prediction method was experimentally evaluated using off-the-shelf IEEE 802.11ad mmWave communication devices and point cloud sensors (i.e., LiDAR and depth camera) in a scenario where mmWave LOS paths were intermittently blocked by pedestrians. 
The experimental results revealed that our point cloud-based method can quantitatively and deterministically predict substantial attenuation of both mmWave RSSI and throughput up to 1000\,ms ahead.
%\appendices
\section*{APPENDIX A. VOXELIZATION ALGORITHM}\label{ap:voxelization}

The detail of the voxelization algorithm is shown in \cref{alg:voxelization}.
First, the min bounds vector $(x_{\min}, y_{\min}, z_{\min})$ and max bounds vector $(x_{\max}, y_{\max}, z_{\max})$ of all points in the point cloud are calculated.
Second, the voxel grid shape $(h, w, d)$ is calculated and the 3D array $V$ representing the voxel grid is initialized with 0.
Finally, for each $n$-th point $\bm{p}_n$, the index $(i_x, i_y, i_z)$ of the corresponding voxel in the voxel grid is calculated and the voxel value is updated to 1.

\begin{algorithm}[t!]
    \caption{Voxelization}
    \label{alg:voxelization}
    \begin{algorithmic}[1]
        \Require {Number of points in point cloud $N$ \vspace{1mm}}
        \Require {Point cloud $\displaystyle \mathcal{P} = \bigcup_{n=0}^{N-1} \left\{\bm{p}_n\right\} $ \vspace{1mm}}
        \Require {Voxel size $s_\mathrm{v} > 0$}
        \Ensure {3D array representing the voxel grid $\bm{V}$}
        \State {$(x_0, y_0, z_0) \leftarrow \bm{p}_0$}
        \State {$(x_{\min}, y_{\min}, z_{\min}) \leftarrow (x_0, y_0, z_0)$}
        \State {$(x_{\max}, y_{\max}, z_{\max}) \leftarrow (x_0, y_0, z_0)$}
        \For{$n$ in $1$ to $N-1$}
            \State {$(x_n, y_n, z_n) \leftarrow \bm{p}_n$}
            \State {$x_{\min} \leftarrow \mathrm{min}(x_{\min}, x_n)$ }
            \State {$y_{\min} \leftarrow \mathrm{min}(y_{\min}, y_n)$}
            \State {$z_{\min} \leftarrow \mathrm{min}(z_{\min}, z_n)$}
            \State {$x_{\max} \leftarrow \mathrm{max}(x_{\max}, x_n)$}
            \State {$y_{\max} \leftarrow \mathrm{max}(y_{\max}, y_n)$}
            \State {$z_{\max} \leftarrow \mathrm{max}(z_{\max}, z_n)$}
        \EndFor \vspace{1mm}
        \State {$\displaystyle h \leftarrow \left\lceil \frac{x_{\max} - x_{\min}}{s_\mathrm{v}}\right\rceil$ \vspace{1mm}}
        \State {$\displaystyle w \leftarrow \left\lceil \frac{y_{\max} - y_{\min}}{s_\mathrm{v}}\right\rceil$ \vspace{1mm}}
        \State {$\displaystyle d \leftarrow \left\lceil \frac{z_{\max} - z_{\min}}{s_\mathrm{v}}\right\rceil$ \vspace{1mm}}
        \State {$\bm{V} \leftarrow \mathbf{bool}[h][w][d]$}
        \For{$i_x$ in $0$ to $h-1$}
            \For{$i_y$ in $0$ to $w-1$}
                \For{$i_z$ in $0$ to $d-1$}
                    \State {$\bm{V}[i_x][i_y][i_z] \leftarrow 0$}
                \EndFor
            \EndFor
        \EndFor
        \For{$n$ in $0$ to $N-1$}
            \State {$\displaystyle (x_n, y_n, z_n) \leftarrow p_n$ \vspace{1mm}}
            \State {$\displaystyle i_x \leftarrow \left\lfloor \frac{x_n - x_{\min}}{s_\mathrm{v}}\right\rfloor$ \vspace{1mm}}
            \State {$\displaystyle i_y \leftarrow \left\lfloor \frac{y_n - y_{\min}}{s_\mathrm{v}}\right\rfloor$ \vspace{1mm}}
            \State {$\displaystyle i_z \leftarrow \left\lfloor \frac{z_n - z_{\min}}{s_\mathrm{v}}\right\rfloor$ \vspace{1mm}}
            \State {$V[i_x][i_y][i_z] \leftarrow 1$}
        \EndFor \\
        \Return {$\bm{V}$}
    \end{algorithmic}
    \Comment{$\lfloor x \rfloor$ represents the greatest integer less than or equal to $x$.}\\
    \Comment{$\lceil x \rceil$ represents the least integer greater than or equal to $x$.}
\end{algorithm}
\section*{APPENDIX B. CONVERSION FROM DEPTH IMAGE TO NORMALIZED POINT CLOUD}\label{ap:dtop}
The conversion procedure from a depth image to a normalized point cloud is shown in \cref{alg:dtop}.
In this paper, this normalized point cloud is also referred to as a depth camera point cloud.
Let $\bm{M}$ be a two-dimensional array representing a depth image.
Let $U$ and $V$ be the width and height of the depth image, respectively, and let $[0, D)$ be the range of depth value $d$.
In the depth image used in this study, $(U, V, D) = (512, 424, 256)$.
A point $\bm{p}$ is assigned in the 3D Cartesian coordinate to each pixel in the depth image.
This process fits all $UV$ points into the $[0,D)^3$ cubic region. 
All depth camera point clouds have 217,088 points in this study.
The aforementioned method is applied to all the depth images to generate depth camera point clouds.
This method can convert from depth images to point clouds without dependence of the camera's intrinsic parameters.
\begin{algorithm}[t!]
    \caption{Conversion from depth image to normalized point cloud}
    \label{alg:dtop}
    \begin{algorithmic}[1]
        \Require {Depth image $\bm{M}$}
        \Require {Width of the depth image $U > 0$}
        \Require {Height of the depth image $V > 0$}
        \Require {Maximum value of depth $D > 0$}
        \Ensure {Point cloud $\mathcal{P}$}
        \State {$\mathcal{P} \leftarrow \varnothing$}\vspace{2mm}
        \State {$c_\mathrm{u} \gets \displaystyle \frac{U-1}{2}$} \vspace{2mm}
        \State {$c_\mathrm{v} \gets \displaystyle \frac{V-1}{2}$} \vspace{2mm}
        \For{$u$ in $0$ to $U-1$}
            \For{$v$ in $0$ to $V-1$}
                \State {$d \leftarrow \bm{M}[u][v]$} \vspace{2mm}
                \State {$x \leftarrow \displaystyle \frac{u-c_\mathrm{u}}{U-1}d + \frac{D}{2}$} \vspace{2mm}
                \State {$y \leftarrow \displaystyle \frac{v-c_\mathrm{v}}{V-1}d + \frac{D}{2}$} \vspace{2mm}
                \State {$z \leftarrow d$}
                \State {$\bm{p} \leftarrow (x,y,z)$}
                \State {$\mathcal{P} \leftarrow \mathcal{P} \cup \{\bm{p}\}$}
            \EndFor
        \EndFor \\
        \Return $\mathcal{P}$
    \end{algorithmic}
\end{algorithm}

\bibliographystyle{IEEEtran}
\bibliography{main}

\begin{IEEEbiography}[{\includegraphics[width=1in,height=1.25in,clip,keepaspectratio]{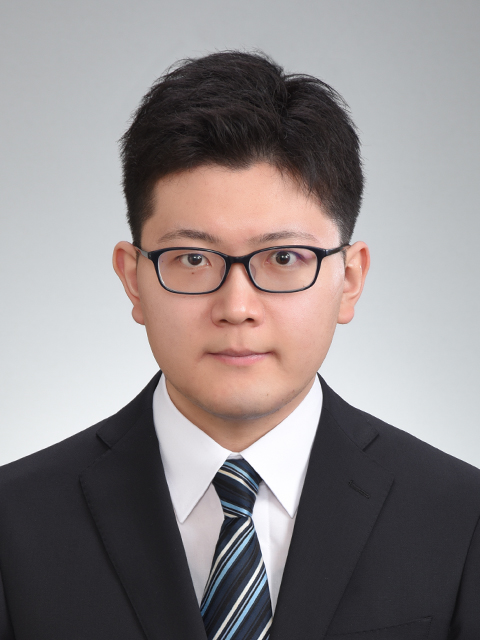}}]{Shoki Ohta}~received the B.E.\ degree in information and communications engineering from Tokyo Institute of Technology in 2022.
He is currently studying toward the M.E.\ degree at the School of Engineering, Tokyo Institute of Technology.
He received the IEEE Vehicular Technology Society (VTS) Japan Young Researcher’s Encouragement Award and Outstanding Student Award from the Department of Information and Communications Engineering, Tokyo Institute of Technology, in 2022.
He is a student member of IEEE and IEICE.
\end{IEEEbiography}

\begin{IEEEbiography}
[{\includegraphics[width=1in,height=1.25in,clip,keepaspectratio]{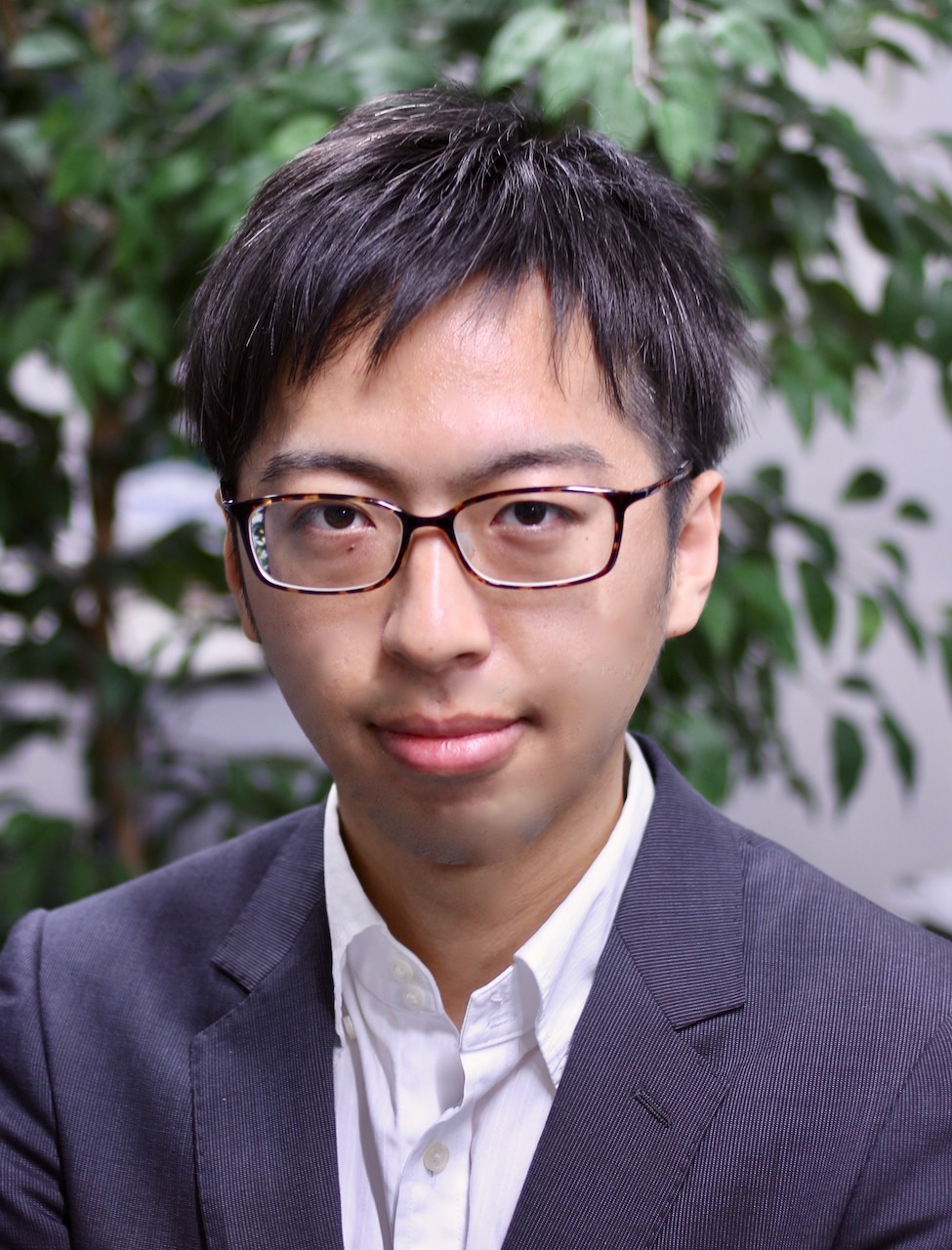}}]{Takayuki Nishio}~(S'11-M'14-SM'20) received the B.E.\ degree in electrical and electronic engineering and the master's and Ph.D.\ degrees in informatics from Kyoto University in 2010, 2012, and 2013, respectively. He had been an assistant professor in the Graduate School of Informatics, Kyoto University from 2013 to 2020. From 2016 to 2017, he was a visiting researcher in Wireless Information Network Laboratory (WINLAB), Rutgers University, United States. He has been an associate professor in the School of Engineering, Tokyo Institute of Technology, Japan, since 2020. His current research interests include machine learning-based network control, machine learning in wireless networks, and heterogeneous resource management.
\end{IEEEbiography}

\begin{IEEEbiography}[{\includegraphics[width=1in,height=1.25in,clip,keepaspectratio]{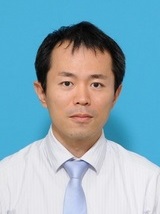}}]{Riichi Kudo}~received the B.S. and M.S.\ degrees in geophysics from Tohoku University, Japan, in 2001 and 2003, respectively. He received the Ph.D.\ degree in informatics from Kyoto University in 2010. In 2003, he joined NTT Network Innovation Laboratories, Japan. He was a visiting fellow at the Center for Communications Research (CCR), Bristol University, UK, from 2012 to 2013, and worked for NTT DOCOMO from 2015 to 2018. He is now working for NTT Network Innovation Laboratories. He received the Young Engineer Award from IEICE and IEEE AP-S Japan Chapter Young Engineer Award in 2006 and 2010, respectively. He is a member of IEICE and IEEE. 
\end{IEEEbiography}

\begin{IEEEbiography}
[{\includegraphics[width=1in,height=1.25in,clip,keepaspectratio]{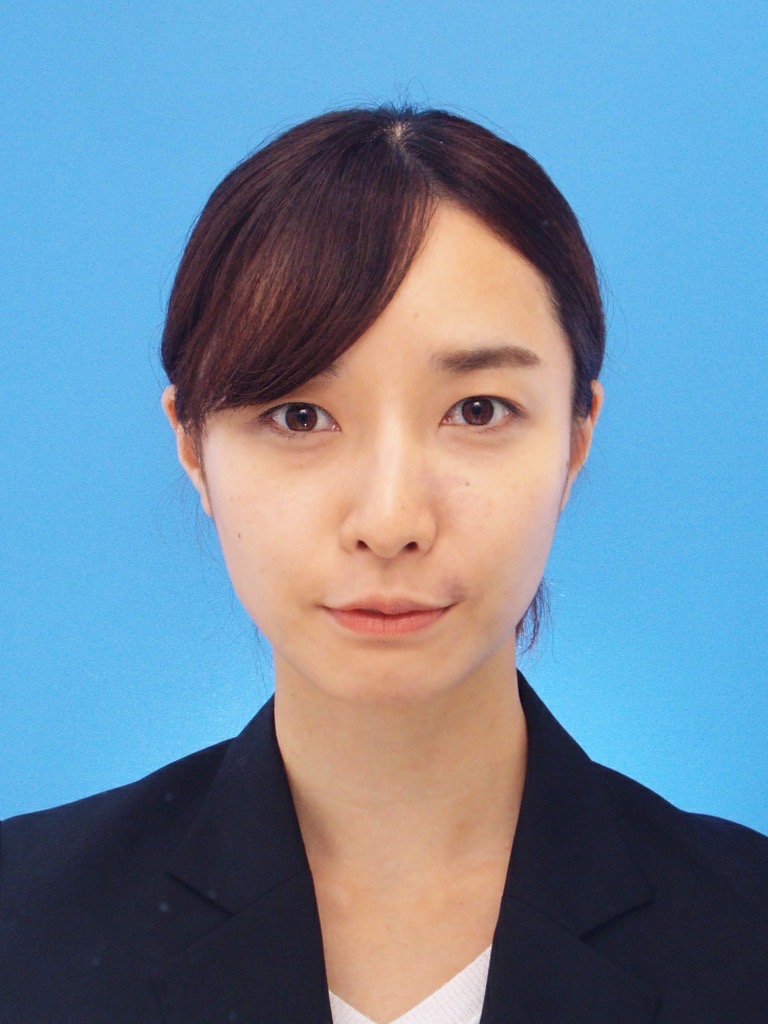}}]{Kahoko Takahashi}~received the B.S.\ degree in seismology and M.S.\ degree in Sensory Information Science from Yokohama City University, Japan, in 2017 and 2019, respectively. In 2019, she joined NTT Network Innovation Laboratories, Yokosuka, Japan. Her current research interests include machine learning. She is a member of IEICE.
\end{IEEEbiography}

\begin{IEEEbiography}
[{\includegraphics[width=1in,height=1.25in,clip,keepaspectratio]{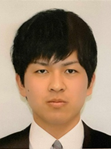}}]{Hisashi Nagata}~received the B.S. and M.S.\ degrees in physics from Tokyo University of Science, Japan, in 2012 and Osaka University, Japan, in 2014, respectively. In 2014, he joined NTT Network Innovation Laboratories, Japan. He worked for NTT WEST from 2017 to 2021. He is now working for NTT Network Innovation Laboratories. He is a member of IEICE.
\end{IEEEbiography}

\end{document}